\documentclass[preprint,aps,12pt,preprintnumbers,eqsecnum,nofootinbib]{revtex4}
\usepackage{xcolor}
\usepackage{graphicx}
\usepackage{epstopdf} 
\usepackage{braket}
\usepackage{amssymb,amsmath,amsfonts,comment,latexsym,graphicx,epstopdf,float}
\usepackage{color}
\usepackage{fancyvrb}
\usepackage{chngcntr}
\usepackage{etoolbox}
\usepackage{ulem}
\usepackage{array}
\usepackage{slashed}

\usepackage[colorlinks=true,citecolor=blue,linkcolor=red,filecolor=cyan,urlcolor=magenta]{hyperref}

\usepackage{float}

\newcommand{\be}{\begin{equation}}
\newcommand{\ee}{\end{equation}}
\newcommand{\ba}{\begin{eqnarray}}
\newcommand{\ea}{\end{eqnarray}}

\newcommand{\grts}{\raise.3ex\hbox{$>$\kern-.75em\lower1ex\hbox{$\sim$}}}
\newcommand{\lets}{\raise.3ex\hbox{$<$\kern-.75em\lower1ex\hbox{$\sim$}}}

\newcommand{\dd}{\text{d}}

{\catcode`\|=\active\gdef\Braket#1{\left<\mathcode`\|"8000\let|\bravert 
{#1}\right>}}

\def\bravert{\egroup\,\vrule\,\bgroup}

\usepackage{color}
\usepackage{amssymb,amsmath,amsfonts}
\usepackage{epstopdf} 
\usepackage{braket}

\usepackage{autobreak}

\begin{document}
%
%
\title{\vspace*{0.5in} 
Asymptotic safety and gauged baryon number
\vskip 0.1in}
\author{Jens Boos}\email[]{jboos@wm.edu}
\author{Christopher D. Carone}\email[]{cdcaro@wm.edu}
\author{Noah L. Donald}\email[]{nldonald@wm.edu}
\author{Mikkie R. Musser}\email[]{mrmusser@wm.edu}
\affiliation{High Energy Theory Group, Department of Physics,
William \& Mary, Williamsburg, VA 23187-8795, USA} 
%
%
\date{June 9, 2022}
\begin{abstract}
We consider a model with gauged baryon number that may be rendered asymptotically safe when gravitational effects above the Planck scale
are taken into account.   We study the ultraviolet fixed points in this theory and determine the restrictions on the parameter space 
of the model at the TeV scale following from the requirement that the asymptotic fixed points are reached. Assuming that the new gauge 
symmetry is broken at the TeV scale, we comment on the phenomenological implications of these restrictions.
\end{abstract}

\maketitle

\section{Introduction} \label{sec:intro}

Extensions of the standard model typically involve a set of new couplings that are only partially constrained by 
low-energy experimental observables.  A well-motivated restriction on the ultraviolet (UV) limit of such a theory is useful when it 
can remove some of this arbitrariness, leading to a more predictive low-energy theory.   In this paper, we consider a 
$Z^\prime$ model whose phenomenology is affected in a meaningful way by the requirement that the theory remain 
asymptotically safe when extrapolated to infinitely high energy scales.  We determine the UV restrictions on the model's 
parameter space at the TeV scale and comment briefly on the phenomenological consequences.

Asymptotic safety was originally proposed by Weinberg in the context of quantum gravity~\cite{weinberg00}; for other related work, see Refs.~\cite{Berges:2000ew,Niedermaier:2006wt}.   If Einstein gravity is treated as a 
quantum field theory, it is well known that the theory is nonrenormalizable, requiring an infinite number of couplings.  In the asymptotic 
safety paradigm, one considers a theory to be unphysical if it includes couplings that flow to a Landau pole at some finite 
energy scale.   Physically acceptable theories correspond to a subspace of the original space of 
parameters called the ultraviolet critical surface.  If this surface is finite-dimensional, then the theory may be rendered predictive 
even though it is nonrenormalizable. Under the renormalization group (RG) flow, couplings on the UV critical surface may run to nontrivial, interacting ultraviolet 
fixed points, or vanishing, ``Gaussian'' fixed points.   We will refer to the theory as asymptotically free if all couplings run to 
Gaussian fixed points, and asymptotically safe if one or more couplings approach an interacting UV fixed point while the rest 
flow to zero.  For a review of asymptotic safety and a comprehensive list of references see, for example, Ref.~\cite{Percacci:2011fr}.

Just as asymptotic safety can reduce the otherwise infinite parameter space of a nonrenormalizable theory, it can 
reduce the finite-dimensional parameter space of a renormalizable one.   This fact has been used to constrain 
standard model extensions in several examples discussed in the recent 
literature~\cite{Reichert:2019car,Hiller:2019mou,Kowalska:2020zve,Domenech:2020yjf,Grabowski:2018fjj,Kwapisz:2019wrl,Hiller:2020fbu,Wang:2015sxe}, 
focusing on issues including dark matter~\cite{Reichert:2019car,Kowalska:2020zve}, the current discrepancy between the standard model 
prediction and the measured value of the muon anomalous magnetic moment~\cite{Hiller:2019mou,Kowalska:2020zve}, various aspects of
neutrino and Higgs sector physics~\cite{Domenech:2020yjf,Grabowski:2018fjj,Kwapisz:2019wrl}, flavor physics~\cite{Bause:2021prv} and 
collider phenomenology~\cite{Hiller:2020fbu,Wang:2015sxe}.   The present work considers another application, adding to this body of literature.   The possibility that baryon number could be gauged has been 
discussed extensively in the past~\cite{Carone:1994aa,Carone:1995pu,FileviezPerez:2010gw,Duerr:2013dza,Duerr:2013lka,FileviezPerez:2014lnj,FileviezPerez:2019jju}, both as a possible way of assuring proton 
stability and for its interesting TeV-scale phenomenology; the latter motivation is relevant for the present work.  The phenomenology 
of the new U(1) gauge boson is largely determined by the gauge coupling $g_B$, the gauge boson mass $m_B$, and a 
parameter $\epsilon$ (defined later) that specifies the kinetic mixing between the U(1)$_B$ and hypercharge gauge groups.   Notably, the U(1)$_B$ 
gauge boson would be entirely leptophobic if not for the kinetic mixing.  Hence, decay channels that may be easier to discern in light of large QCD 
backgrounds (\textit{i.e.}, decays to charged dileptons rather than dijets) are entirely controlled by the undetermined kinetic mixing 
parameter.   The same parameter also controls mixing of the U(1)$_B$ and electroweak gauge bosons that is crucial in determining
the constraints from precise electroweak measurements at the $Z$-boson scale.   One way of fixing the kinetic mixing parameter, discussed in
Ref.~\cite{Carone:1995pu}, is to require that it vanishes at some scale by embedding the two Abelian gauge group factors into a non-Abelian one.  
Here we explore a more economical alternative---and one motivated by the eventual inclusion of gravity---that asymptotic safety allows us
to predict the kinetic mixing in terms of other model parameters based on the requirement that appropriate fixed points are reached in the UV.  
This reduces the space of possibilities for the properties of the $Z^\prime$ boson, and provides a guide for discerning the model at 
collider experiments.\footnote{We note that Ref.~\cite{Wang:2015sxe} also considers asymptotic safety in a leptophobic model, but 
one in which only third-generation quarks are charged under a new U(1).  This implies a different fixed point structure than the one predicted
by the model proposed here.  More significantly, the model in Ref.~\cite{Wang:2015sxe} has a serious problem: the stated charge assignments for the fermions that are vector-like under the standard model gauge group forbid the only possible Yukawa couplings that could generate their masses. This leads to massless, electrically charged fermions, ruling 
out the model.}

Our paper is organized as follows. In Sec.~\ref{sec:two} we define the model. In Sec.~\ref{sec:three}, we study the UV behavior of the theory, 
identifying a number of scenarios where the model can be extrapolated to infinite energy with some couplings reaching nontrivial fixed points, and 
where vacuum stability is maintained.  We determine what these UV boundary conditions imply about the allowed parameter values at the TeV scale.
In Sec.~\ref{sec:four}, we comment on the phenomenological implications of these results, presenting one example in which the branching fraction of 
the $Z^\prime$ boson into standard model particles is predicted.  We summarize our conclusions in Sec.~\ref{sec:conc}.  Appendix \ref{app} contains the complete list of 
one-loop renormalization group $\beta$-functions used in our analysis.

\section{Model} \label{sec:two}

We consider an extension of the standard model in which baryon number, U(1)$_B$, is gauged.  We normalize that gauge coupling so that that baryon number charge
is $+1$ for a proton or neutron;  the baryon number is $0$ for any standard model lepton.  A right-handed neutrino is included so that Dirac neutrino masses are possible.   
We do not address the problem of the flavor structure of the standard model, nor the smallness of neutrino masses, but remain content with the fact that the allowed 
couplings of the model are sufficient to accommodate all observed fermion masses and mixing angles.   The charges for the particle content described thus far are 
shown in Table~\ref{table:one}.
\begin{table}[!h]
\begin{tabular}{ccccccccc} \hline\hline
&  \hspace{1em} & SU(3)$_C$ & \hspace{1em} & SU(2)$_W$ &  \hspace{1em} &U(1)$_Y$ & \hspace{1em}  &U(1)$_B$ \\ \hline
 $q_L$ & &$3$ && $2$ && $1/6$ && $1/3$ \\ 
 $u_R$ & &$3$ && $1$ && $2/3$ && $1/3$ \\ 
 $d_R$ & &$3$ && $1$ && $-1/3$ && $1/3$ \\ 
 $\ell_L$ & &$1$ && $2$ && $-1/2$ && $0$ \\ 
 $e_R$ & &$1$ && $1$ && $-1$ && $0$ \\ 
 $\nu_R$ & &$1$ && $1$ && $0$ && $0$ \\ \hline\hline
 \end{tabular}
 \caption{Charge assignments for a single generation of standard model fields, including a right-handed neutrino.}  \label{table:one}
 \end{table}

Additional fermions must be added to assure the cancellation of gauge and gravitational anomalies.   We assume three generations of  Dirac fermions $\psi^\ell$, $\psi^e$ and $\psi^\nu$ 
that are vector-like under the standard model gauge group, with quantum numbers identical to those of the lepton fields $\ell_L$, $e_R$ and $\nu_R$, respectively; the new 
fields are chiral under U(1)$_B$.  We temporarily denote the U(1)$_B$ charges of  $\psi^\ell_{L,R}$, $\psi^e_{L,R}$ and $\psi^\nu_{L,R}$ as $x_{L,R}$,  $y_{L,R}$, and $z_{L,R}$, respectively.
For simplicity, we seek the cancellation of anomalies within each generation separately.  The anomaly cancellation constraints are then summarized as follows:
\begin{table}[b]
\begin{tabular}{ccccccccc} \hline\hline
&  \hspace{1em} & SU(3)$_C$ & \hspace{1em} & SU(2)$_W$ &  \hspace{1em} &U(1)$_Y$ & \hspace{1em}  &U(1)$_B$ \\ \hline
 $\psi^\ell_L$ & &$1$ && $2$ && $-1/2$ && $x_L =0$\\ 
 $\psi^\ell_R$ & &$1$ && $2$ && $-1/2$ && $x_R =1$ \\ 
 $\psi^e_L$ & &$1$ && $1$ && $-1$ && $y_L =1$ \\ 
 $\psi^e_R$ & &$1$ && $1$ && $-1$ && $y_R =0 $ \\ 
 $\psi^\nu_L$ & &$1$ && $1$ && $0$ && $z_L =1$ \\ 
 $\psi^\nu_R$ & &$1$ && $1$ && $0$ && $z_R =0$ \\ \hline\hline
 \end{tabular}
 \caption{Charge assignments for the vector-like fields, for a single generation.  The last column shows the anomaly-free solution 
 discussed in the text.}\label{table:two}
 \end{table}
\begin{itemize}
\item  U(1)$_B$ SU(3)$^2$: This anomaly is proportional to $2 \cdot \frac{1}{3} - \frac{1}{3}-\frac{1}{3} =0$, and vanishes without help from the vector-like sector.
\item  U(1)$_B$ SU(2)$^2$: This anomaly is proportional to $3 \cdot \frac{1}{3} + x_L - x_R$, which implies
\begin{equation}
x_L-x_R=-1 \,\,\,.
\label{eq:xlxr}
\end{equation}
\item U(1)$_B$ U(1)$_Y^2$: This anomaly is proportional to $-\frac{1}{2}+\frac{1}{2}(x_L-x_R)+(y_L-y_R)$. With the constraint of Eq.~(\ref{eq:xlxr}), this implies
\begin{equation}
y_L-y_R=+1
\label{eq:ylyr}
\end{equation}
\item
U(1)$_B^2$ U(1)$_Y$: The standard model particles do not contribute to this anomaly, but the new particles do, so that
\begin{equation}
-(x_L^2-x_R^2) - (y_L^2-y_R^2)=0 \,\,\, .
\label{eq:B2Y}
\end{equation}
\item U(1)$_B^3$:  Again, there is no contribution in total from the standard model particles, but the new particles contribute:
\begin{equation}
2\,(x_L^3-x_R^3)+(y_L^3-y_R^3)+(z_L^3-z_R^3)=0 \,\,\, .
\end{equation}
\item gg U(1)$_Y$:  Here, $g$ refers to a graviton.  The hypercharge gravitational anomaly cancels in the standard model, and this is not affected by the new particles which are vector-like in their standard model charges.  
\item gg U(1)$_B$:   In this case, the anomaly is proportional to $2\, (x_L-x_R)+(y_L-y_R)+(z_L-z_R)$.   With the constraints of Eqs.~(\ref{eq:xlxr}) and (\ref{eq:ylyr}), this implies
\begin{equation}
z_L-z_R = +1 \,\,\, .
\end{equation}
\end{itemize}
All the constraints are satisfied with the choice $x_R=y_L=z_L=+1$ and $x_L=y_R=z_R=0$, as indicated in Table~\ref{table:two}.

We assume that the U(1)$_B$ symmetry is spontaneously broken by a complex scalar field $\phi$ which has baryon number $+1$ and is a singlet under the standard model gauge group.   The
charge assignment of $\phi$ is fixed by the requirement that it allows Yukawa couplings which generate masses for the $\psi$ fields when $\phi$ develops a vacuum expectation value (vev).   One finds that the desired
Yukawa couplings are given by
\begin{equation}
{\cal L}_{y} =  \overline{\psi^\ell_L} \, y_1 \, \psi^\ell_R \, \phi^* + \overline{\psi^e_L} \, y_2 \, \psi^e_R \, \phi  + \overline{\psi^\nu_L} \, y_3 \, \psi^\nu_R \, \phi + \mbox{ H.c.}  \,\, ,
\label{eq:yuky}
\end{equation}
where the $y_i$ are three-by-three matrices.\footnote{For simplicity, we omit possible Majorana masses for $\nu_R$ and $\psi^\nu_R$.}
 Since there are pairs of fields that have identical quantum numbers, namely ($\psi^\ell_L$, $\ell_L$), ($\psi^e_R$, $e_R$) and 
($\psi^\nu_R$, $\nu_R$), we may choose a field basis in which there are no Yukawa couplings involving $\phi$ that mix a heavy and light field, such as $ \overline{\ell_L} \psi^\ell_R\phi^*$.
However, heavy-light couplings are possible involving the standard model Higgs field $H$:
\begin{equation}
{\cal L}_{\kappa} = \overline{\psi^\ell_L} \, H \, \kappa_1 e_R + \overline{\ell_L} \, H \, \kappa_2 \, \psi^e_R + \overline{\ell_L} \, \widetilde{H} \, \kappa_3 \psi^\nu_R + \mbox{ H.c.} \,\, ,
\label{eq:yukk}
\end{equation}
where the $\kappa_i$ are also three-by-three matrices. This set of Yukawa couplings serves a useful purpose phenomenologically, as it assures that the heavy fields 
can decay to light fields, thereby avoiding unwanted stable charged particles. (For the stringent bounds on heavy, stable charged particles, see \textit{Other Particle Searches} in Ref.~\cite{ParticleDataGroup:2020ssz}). 
 
It is worth noting that the spontaneous breaking of baryon number through the $\phi$ vev does not lead to any problems with proton 
decay.  The Lagrangian has an anomalous global U(1) baryon number symmetry acting exclusively on the quark fields $q$, $u$ and $d$, even in the 
presence of a $\phi$ vev. This implies that any gauge-invariant, dimension-six operator that contributes to proton decay, and violates this global 
symmetry, cannot be generated at any order in perturbation theory, where it might only be suppressed by a mass scales appearing in the Lagrangian.   
On the other hand, Planck-suppressed dimension-six operators, if present, would be sufficiently suppressed as the lower bound on the scale of 
dimension-six operators from proton decay is typically ${\cal O}(10^{16})$~GeV~\cite{ParticleDataGroup:2020ssz}.  Interestingly, there is some 
evidence that asymptotically safe gravity may preserve global symmetries, in which case even these operators would not arise~\cite{Eichhorn:2017eht};
for additional discussion, see Ref.~\cite{Kowalska:2020zve}.
 
The rest of the theory consists of the scalar sector
\begin{equation}
V(\phi,H) = -m_H^2 H^\dagger H + \frac{\lambda}{2} (H^\dagger H)^2 
-m_\phi^2 \phi^* \phi + \frac{\lambda_\phi}{2} (\phi^* \phi)^2 
+\lambda_m \phi^* \phi H^\dagger H \,\,\, ,
\label{eq:thepot}
\end{equation}
which involves the new couplings $\lambda_m$ and $\lambda_\phi$, and the gauge kinetic mixing between U(1)$_B$ and hypercharge
\begin{equation}
{\cal L} \supset -\frac{1}{4} B^{\mu\nu} B_{\mu\nu} -\frac{1}{4} F^{\mu\nu}_Y F_{\mu\nu}^Y +\frac{\epsilon}{2} B^{\mu\nu} F_{\mu\nu}^Y\,\,\, ,
\label{eq:u1kin1}
\end{equation}
which involves the kinetic mixing parameter $\epsilon$.  Thus, in addition to the parameters of the standard model, the theory we have just defined has one new gauge coupling 
$g_B$, a gauge-kinetic mixing parameter $\epsilon$, two new Higgs sector couplings $\lambda_\phi$ and $\lambda_m$, and the new Yukawa couplings in 
Eqs.~(\ref{eq:yuky}) and (\ref{eq:yukk}).   If one temporarily rescales the gauge fields so that the gauge couplings appear in the kinetic terms, Eq.~(\ref{eq:u1kin1}) takes the form
\begin{equation}
{\cal L} = - \frac{1}{4} (G^{-2})_{AB} F^A_{\mu\nu} F^{B\,\mu\nu}  \,\,\, ,
\label{eq:G2}
\end{equation}
where the indices run over the two-dimensional space of Abelian gauge fields.
In studying the renormalization group equations (RGEs) for models of this type, it is conventional~\cite{pyrate} to redefine the gauge field 
basis so that the matrix $G$ has the upper-triangular form
\begin{equation}
G = \left(\begin{array}{cc} g_Y & \displaystyle \frac{\epsilon}{\sqrt{1-\epsilon^2}} \, g_Y \, \\[15pt] 0 & \displaystyle \frac{1}{\sqrt{1-\epsilon^2}} \, {g_{B}}_0 \end{array}\right) \equiv \left(\begin{array}{cc} g_Y & 
\tilde{g} \\ 0 & g_B 
\end{array}\right) \,\,\,.
\label{eq:uptriang}
\end{equation}
Here ${g_{B}}_0$ is the baryon number gauge coupling in the original basis. The RGEs are then conveniently expressed in terms of $\tilde{g}$, 
$g_Y$ and $g_B$.  In the basis where the kinetic terms are canonical in form, the covariant derivative on a generic field $\chi$ may be expressed as
\begin{equation}
D_\mu \chi = [\partial_\mu  - i \, g_B B_\mu Q_B -i \, (g_Y A^Y_\mu + \tilde{g} \, B_\mu) Q_Y ]\, \chi \, ,
\label{eq:covd}
\end{equation}
where $Q_B$ and $Q_Y$ are the baryon number and hypercharges of $\chi$, respectively.  This is a convenient form for studying some of the phenomenological consequences of the RGE output.

Finally, following the conventional approach
~\cite{Reichert:2019car,Kowalska:2020zve,Domenech:2020yjf,Grabowski:2018fjj,Kwapisz:2019wrl,Hiller:2020fbu,Wang:2015sxe}, we adopt a simplified flavor structure of the theory for use in our numerical RGE analysis: we ignore standard model lepton Yukawa couplings and assume that the matrices $y_i$ and $\kappa_i$, $i=1 \ldots 3$, are each proportional to the identity matrix;  in other words, these couplings will be taken to represent six parameters rather than six arbitrary matrices.  These assumptions are consistent with the renormalization group running of the couplings for the following reason:  In the absence of standard model lepton Yukawa couplings, the Lagrangian has a U(3)$_\ell \times$U(3)$_e$ chiral symmetry.    Writing the charge assignments via the representation
pair $({\bf r}_\ell,{\bf r}_e)$, this symmetry can be extended to the heavy leptons,
\begin{align}
& e_R \sim \psi_L^\ell \sim \psi_R^\ell \sim ({\bf 1},{\bf 3})  \nonumber \\
&\ell_L \sim \psi^e_L \sim  \psi^e_R \sim \psi^\nu_L \sim  \psi^\nu_R \sim ({\bf 3},{\bf 1}) \,\,\, ,
\end{align}
provided that the $y_i$ and $\kappa_i$  are  proportional to three-by-three identity matrices.  This global symmetry is not broken by any perturbative interaction, allowing us to conclude that the simple flavor structure \textit{ansatz} we have assumed will not be altered by RGE running.

\section{Fixed point analysis} \label{sec:three}

With the model now fully developed, let us consider the structure of its renormalization group flow in detail. In particular, we will search for fixed points at one loop in perturbation theory, with gravitational effects 
parametrized, and consider in which parts of parameter space this model exhibits asymptotic safety. We extract the $\beta$-functions using PyR@TE 3~\cite{pyrate}; see also appendix~\ref{app}.

\subsection{Generalities}

Let us briefly recall some terminology. Consider some couplings $g_i$ with their associated $\beta$-function denoted as $\beta_i$. To study the behavior of the RG flow around a fixed point $g_{i\star}$, consider the expansion
\begin{align}
\label{eq:fp-expansion-1}
\beta_i &= M_i{}^j \, \delta_j + P_i{}^{jk} \, \delta_j \, \delta_k + \mathcal{O}\left(\delta^3\right) \, ,
\end{align}
where $\delta_j \equiv g_j - g_{j\star}$ and  we defined the coefficients
\begin{align}
M_i{}^j &\equiv \left.\frac{\partial \beta_i}{\partial g{}_j}\right|_\star \, , \qquad
P_i{}^{jk} \equiv \frac12 \left.\frac{\partial^2 \beta_i}{\partial g{}_j \partial g{}_k}\right|_\star \, .
\end{align}
The matrix $M{}_i{}^j$ has eigenvectors $v_j{}^k$, where $k$ labels the vectors such that $M{}_i{}^j v{}_j{}^k = \vartheta{}_k v{}_i{}^k$ (no summation over $k$). At linear order in $\delta_i$, Eq.~\eqref{eq:fp-expansion-1} is solved by
\begin{align}
g_i(\mu) = g_{i\star} + \sum\limits_k c_k v{}_i{}^k \left(\frac{\mu}{\Lambda}\right)^{\vartheta_k} \, ,
\end{align}
where $\Lambda$ is an arbitrary reference energy scale defining the origin of ``renormalization time'' $t = \ln(\mu/\Lambda)$. The $c_k$ are subject to the constraint $g_i(\mu\rightarrow\infty) = g_{i\star}$, which 
requires that $c_k \equiv 0$ for all $k$ with $\vartheta_k \geq 0$. The eigendirections $v{}_i{}^k$ in coupling space are classified according to 
the sign of their respective eigenvalues: $\vartheta_k < 0$ (``relevant''), $\vartheta_k = 0$ (``marginal''), and $\vartheta_k > 0$ (``irrelevant'').  
Consequently, the UV critical surface is spanned by all the relevant eigendirections, as well as any marginal ones that lead to flow towards the fixed point.  The latter behavior in the case of marginal directions, however, cannot be established by considering only the linear terms in Eq.~(\ref{eq:fp-expansion-1}), but requires study of the $\beta$-functions at higher order.

It is generally assumed that the influence of gravity can be safely neglected when considering particle physics well below the Planck scale.  However, since the renormalization group flow extends to infinite
energies in the scenarios of interest to us, gravitational corrections to the $\beta$-functions at and above the Planck scale need to be taken into account. 
(For a different approach towards realizing asymptotic safety see, for example, Ref.~\cite{Pelaggi:2017abg}.)
The precise form of these corrections depends on the exact matter content and gravitational theory under consideration, and they have been computed in several scenarios in the so-called Einstein--Hilbert truncation~\cite{Reuter:1996cp}; see also Refs.~\cite{Toms:2010vy,Harst:2011zx,Christiansen:2017gtg,Eichhorn:2017lry}. In a generic but simplified picture adopted in phenomenological literature \cite{Reichert:2019car,Kowalska:2020zve,Wang:2015sxe}, the gravitational corrections to the $\beta$-functions are modeled by ($M_\text{Pl} = 1.2 \times 10^{16}\,\text{TeV}$)
\begin{align}
\beta(g_i) = \frac{1}{(4\pi)^2}\beta^{(1)}(g_i) -  \theta\left(\mu-M_\text{Pl} \right)  \,f_g  \, g_i \, , \label{eq:gravfs} \\
\beta(y_i) = \frac{1}{(4\pi)^2}\beta^{(1)}(y_i) -  \theta\left(\mu-M_\text{Pl} \right) \, f_y  \, y_i \, , \\
\beta(\lambda_i) = \frac{1}{(4\pi)^2}\beta^{(1)}(\lambda_i) -  \theta\left(\mu-M_\text{Pl} \right) \, f_\lambda \, \lambda_i \, ,
\end{align}
where $\theta$ denotes the Heaviside step function.   Here, for compactness, $g_i$, $y_i$ and $\lambda_i$ represent the sets of couplings 
$\{g_1,g_2,g_3,g_B,\tilde{g} \}$, $\{y_t,y_b,y_1,y_2,y_3,\kappa_1,\kappa_2,\kappa_3 \}$ and $\{\lambda,\lambda_\phi,\lambda_m\}$, respectively.  Note that the universal coupling of gravity to matter implies that these corrections $f_g$, $f_y$, and $f_\lambda$ are universal in the gauge, Yukawa, and quartic sectors of the model, respectively.\footnote{Perhaps a more transparent way to understand the universality of the gravity correction term
in the gauge coupling sector, when kinetic mixing is present, is to write the renormalization group equation in terms of the coupling matrix $G^2_{AB}$, 
defined in Eq.~(\ref{eq:G2}).  Working in this basis, $G^2_{AB}$ encodes all the dependence on the gauge couplings in any diagrammatic calculation. A universal gravitational correction term would be introduced through a term proportional to this coupling matrix,
\begin{align*}
\frac{\dd G_{AB}^2}{\dd t} = \frac{1}{2} \frac{1}{16 \pi^2} \left[G^2_{AC} \beta^{(1)}_{CD} G^2_{DB} + (A\leftrightarrow B)\right] - 2 \, G^2_{AB} \, f_g \,\, , 
\end{align*}
where the form of the non-gravitational part of the RGE can be found in Eq.~(5.1) of Ref.~\cite{Poole:2019kcm}.  This reduces to Eq.~(\ref{eq:gravfs}) when expressed in terms of the component couplings, and yields the desired form for the gravitational corrections in the case where kinetic mixing is vanishing.}  The form of the gravitational correction in Eq.~(\ref{eq:gravfs}) for the gauge couplings was shown first in Ref.~\cite{Folkerts:2011jz} where it was found that $f_g$ is renormalization scheme dependent and either positive or zero; the assumption that $f_g>0$ that we adopt here is standard in the phenomenological literature, and corresponds to schemes that break certain classical gauge-gravity symmetries that would otherwise lead to a vanishing result~\cite{Folkerts:2011jz}.   For other discussion of $f_g$ see Refs.~\cite{Toms:2010vy,Harst:2011zx,Christiansen:2017gtg,Eichhorn:2017lry}.   On the other hand, the signs and magnitudes of the Yukawa and quartic gravitational corrections are typically less constrained. In what follows, we shall treat the triplet $(f_g,f_y,f_\lambda)$ as an input parameter to our model, and explore the fixed point structure for a given triplet.

\subsection{Fixed points and critical surfaces}

It is instructive to first consider the U(1)$_B \times$U(1)$_Y$ sector of our theory, including the kinetic mixing.   Henceforth, we use the SU(5) normalization of hypercharge, $g_1 \equiv \sqrt{5/3} \, g_Y$; then
the gravity-corrected $\beta$-functions are given by
\begin{align}
\beta^{(1)}(g_1) &= +\frac{77}{10} g_1^{3} -  \theta(\mu - M_\text{Pl}) \, \hat{f}_g \, g_1 \, , \label{eq:beta-g1} \\
\beta^{(1)}(g_B) &= +11 g_B^{3} + \frac{77}{6} g_B \tilde{g}^{2} -  \frac{16}{3} g_B^{2} \tilde{g} - \theta(\mu - M_\text{Pl}) \, g_B \, \hat{f}_g \, , \label{eq:beta-gB} \\
\beta^{(1)}(\tilde{g}) &= -\frac{16}{5} g_1^{2} g_B -  \frac{16}{3} g_B \tilde{g}^{2} + \frac{77}{5} g_1^{2} \tilde{g} + 11 g_B^{2} \tilde{g} + \frac{77}{6} \tilde{g}^{3} -  \theta(\mu - M_\text{Pl}) \, \tilde{g} \, \hat{f}_g \, , \label{eq:beta-gt}
\end{align}
where we use the notation $\hat{f}_A \equiv (4\pi)^2\,f_A$ with $A=g,y,\lambda$.

For energy scales $\mu < M_\text{Pl}$, Eq.~\eqref{eq:beta-g1} follows the usual logarithmic running
\begin{equation}
\alpha_1(\mu)^{-1} = \alpha_1(\mu_0)^{-1} - \frac{1}{2 \pi} \frac{77}{10} \ln\left(\frac{\mu}{\mu_0}\right)\, ,\,\,\,\,\, \alpha_1(\mu) \equiv \frac{g_1^2(\mu)}{4 \pi} \, ,
\end{equation}
or equivalently,
\begin{align}
g_1(\mu) = \frac{g_1(\mu_0)}{\sqrt{1 - \frac{77}{10} \frac{2 \, g_1^2(\mu_0)}{(4\pi)^2}\ln\left(\frac{\mu}{\mu_0}\right)}} \, ,
\end{align}
and, if not for the gravity correction $\hat{f}_g$ taking over at $\mu > M_\text{Pl}$, the hypercharge gauge coupling would hit a Landau pole eventually. Note that the $g_1$ fixed point structure is independent of the couplings $g_B$ and $\tilde{g}$ at one loop. In the trans-Planckian regime, the fixed point criterion for $g_1$ reads as follows:
\begin{align}
\left(\frac{77}{10}g_{1\star}^2-\hat{f}_g\right)g_{1\star} = 0 \, .
\end{align}
This equation has a trivial solution, corresponding to a Gaussian fixed point, as well as a nontrivial solution, corresponding to an interacting fixed point.  It is clear that these are the only two fixed point scenarios: if $\hat{f}_g$ is too small, then the cubic term in $\beta^{(1)}(g_1)$ will dominate and drive $g_1$ to infinite values. If $\hat{f}_g$ is at a critical value $\hat{f}_g^\text{crit}$, however, $g_1$ will attain a fixed point exactly at the Planck scale. And if $\hat{f}_g$ is larger than this critical value, the linear term in $\beta^{(1)}(g_1)$ dominates and drives the coupling to zero at infinite energies:
\begin{align}
\begin{split}
\hat{f}_g & < \hat{f}_g^\text{crit}  \hspace{2em} \text{no $g_1$ fixed point} \, , \\
\hat{f}_g & = \hat{f}_g^\text{crit}  \hspace{2em} \text{interacting $g_1$ fixed point} \, , \\
\hat{f}_g & > \hat{f}_g^\text{crit}  \hspace{2em} \text{Gaussian $g_1$ fixed point} \, .
\end{split}
\end{align}
In what follows, we shall discuss the two fixed-point scenarios in more detail.  We present numerical values to five significant figures since in some instances this is relevant for hitting unstable fixed point values when running up from the TeV scale. It is interesting to note that there is evidence that fixed points in the gauge sector do not destroy a nontrivial fixed point for the gravitational coupling~\cite{Christiansen:2017cxa}, which makes the phenomenological approach described here sensible when considered in a broader context. For a review of the effects of matter couplings on asymptotic safety in the gravity sector, see Ref.~\cite{Litim:2011cp}.

\subsubsection{Gaussian $g_1$ fixed point} \label{subsec:gauss1}
The Gaussian fixed point, $g_{1\star} = 0$, is attainable if the gravity correction dominates at the Planck scale, amounting to the condition
\begin{align}
\hat{f}_g > \hat{f}_g^\text{crit} \equiv \frac{77}{10} \, g_1^2(M_\text{Pl}) = \frac{g^2_1(\mu_0)}{\frac{10}{77} - \frac{2 \, g_1^2(\mu_0)}{(4\pi)^2}\ln\left(\frac{M_\text{Pl}}{\mu_0}\right)} \approx 7.9610 \, ,
\label{eq:gausfgval}
\end{align}
where the numerical value follows from $\alpha^{-1}(\mu_0) = 57.527$ at $\mu_0=1$~TeV~\cite{Hiller:2020fbu}.  
Assuming that the above bound is satisfied, the gravity correction will then drive $g_1$ to zero asymptotically such that $g_{1\star} = 0$. Then, Eqs.~\eqref{eq:beta-gB} and \eqref{eq:beta-gt} take the form
\begin{align}
\left.\beta^{(1)}(g_B)\right|_{g_{1\star}=0} &= g_B \, E(g_B, \tilde{g}, \hat{f}_g) \, , \\
\left.\beta^{(1)}(\tilde{g})\right|_{g_{1\star}=0} &= \tilde{g} \, E(g_B, \tilde{g}, \hat{f}_g) \, , \\
E(g_B,\tilde{g},\hat{f}_g) &= \frac{77}{6}\,  \tilde{g}^2 + 11 \,g_B^2 - \frac{16}{3} g_B \, \tilde{g} - \hat{f}_g \, .
\end{align}
The fixed points are given by the Gaussian fixed point $(g_{B\star},\tilde{g}_\star)=(0,0)$, as well as an ellipse $E(g_{B\star},\tilde{g}_\star,\hat{f}_g) = 0$. This ellipse is rotated by an angle $\theta$ in $g_B\tilde{g}$-plane,
\begin{align}
\tan\theta = \frac{32}{11 + \sqrt{1145}} \, , \quad \theta \approx 36^\circ \, .
\end{align}
The new gauge coupling $g_B$ and the mixing $\tilde{g}$ are bounded by
\begin{align}
g_B \in \left[0, \sqrt{\tfrac{231 \hat{f}_g}{2413}} \, \right] \, , \quad
\tilde{g} \in \left[ -\sqrt{\tfrac{6 \hat{f}_g}{77}}, \sqrt{\tfrac{198 \hat{f}_g}{2413}} \, \right] \, .
\end{align}
Note that $\tilde{g}$ can be negative, given its relation to the Lagrangian parameter $\epsilon$ in Eq.~(\ref{eq:uptriang}).  It is obvious from these expressions that in the limiting case of $\hat{f}_g \rightarrow 0$, that is, for vanishing gravity corrections, the ellipse shrinks to zero size and only the Gaussian fixed point survives. In other words, this non-trivial structure is generated by the gravitational corrections.

In order to understand the behavior in the $g_B\tilde{g}$-sector better, consider a graphical visualization of the two $\beta$-functions in Fig.~\ref{fig:1}. As it turns out, the ellipse corresponds to an unstable collection of fixed points, also referred to as ``UV repulsive'' in the asymptotic safety terminology, where all values inside the ellipse flow towards the Gaussian fixed point $(g_{B\star},\tilde{g}_\star)=(0,0)$. All values outside the ellipse flow to infinite values. In other words, the ellipse corresponds to a projection of the UV critical surface into the subspace $g_1=0$. This implies that the values of $g_B$ 
and $\tilde{g}$ are not independent if they are required to reach nontrivial fixed point values in the UV.

Given the UV critical surface, it is now pertinent to determine what range of coupling values at $\mu_0 = 1\,\text{TeV}$ flow to those fixed points, where we have selected a value for $\mu_0$ that is
representative of high-energy collider physics experiments.  Since the UV ellipse of fixed points is unstable, it must be hit exactly when running up from lower energies,  leading to greater predictivity than one 
would obtain in the case of fixed points that are attractive.  There arises a technical complication: Since the Gaussian $g_1$ fixed point is an asymptotic one, attained at infinite energy, it is not possible to fully model this numerically. In order to keep the treatment tractable, we define a large energy scale
\begin{equation}
\ln \left(\frac{\mu_\text{max}}{M_{\rm Pl}}\right) = 100 \,\,\, ,
\end{equation}
or equivalently, $t_\text{max} = \ln(\mu_\text{max}/\mu_0) \approx 137$.   The scale $\mu_\text{max}$ is approximately 43 orders of magnitude higher than the Planck energy, high enough so that 
$g_1(\mu_\text{max}) \ll 1$ when $\hat{f}_g > \hat{f}_g^\text{crit}$.  The choice $f_g = 0.1$, for example, gives $g_1(\mu_\text{max}) \approx 6.5 \times 10^{-5}$.   A linearized analysis of the RG flow near the $g_1$ 
fixed point suggests that we may assume that the values of $g_B$ and $\tilde{g}$ at the same scale are given approximately by their fixed point values on the ellipse 
$E(g_{B\star},\tilde{g}_\star,\hat{f}_g) = 0$, completing our set of  boundary conditions at the high energy scale $\mu_\text{max}$.  This provides us with a method of mapping the UV critical surface to a 
corresponding surface renormalized at $\mu_0 = 1\,\text{TeV}$. Once this surface is 
obtained, we may verify by running up from $\mu_0$, to scales even higher than $\mu_\text{max}$, that the couplings approach the desired fixed point, 
providing a numerical sanity check of our computations. In Fig.~\ref{fig:2}, we show (i) the exact UV critical surface, (ii) the resulting values at the 
Planck scale, and, finally, (iii) the resulting values at $\mu_0 = 1\,\text{TeV}$. 

We conclude that asymptotically safe solutions with a Gaussian $g_1$ fixed point and nontrivial fixed points in the $g_B\tilde{g}$-plane lead to a correlation between the parameters $g_B$ and $\tilde{g}$ 
at low energy scales.

\subsubsection{Interacting $g_1$ fixed point} \label{subsec:nontrivial1}
\begin{figure}[!htb]
\centering
\includegraphics[width=0.48\textwidth]{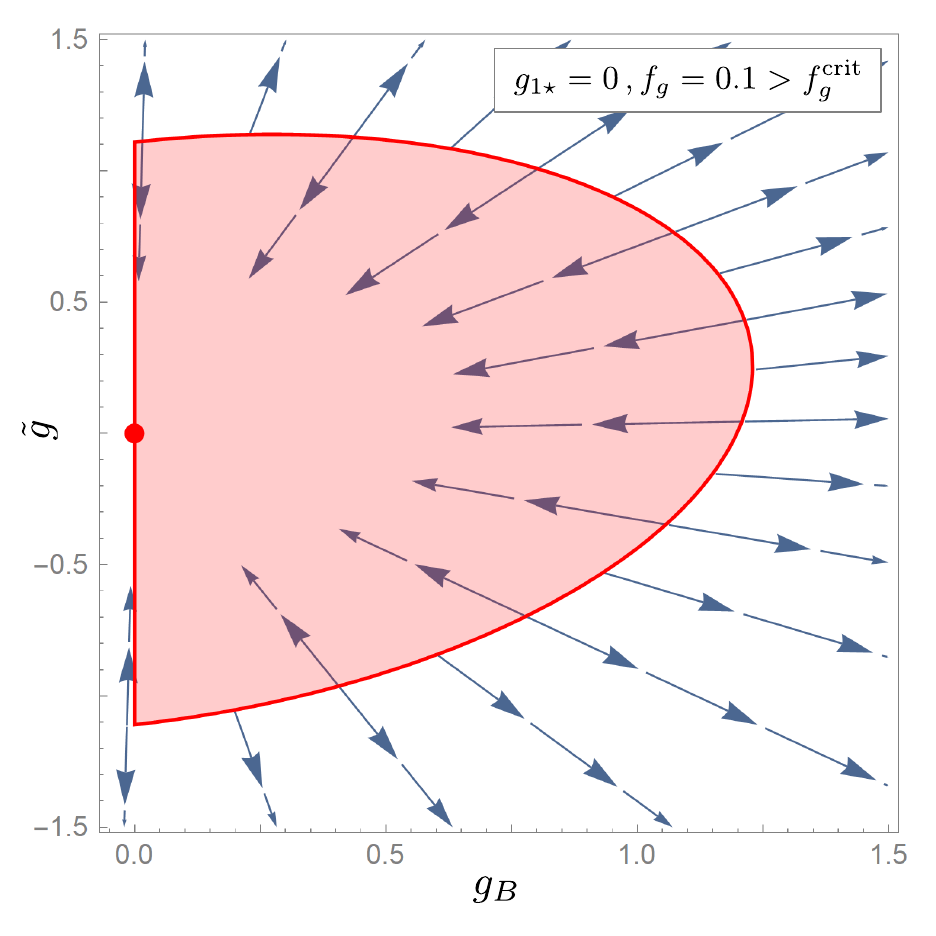}
\includegraphics[width=0.48\textwidth]{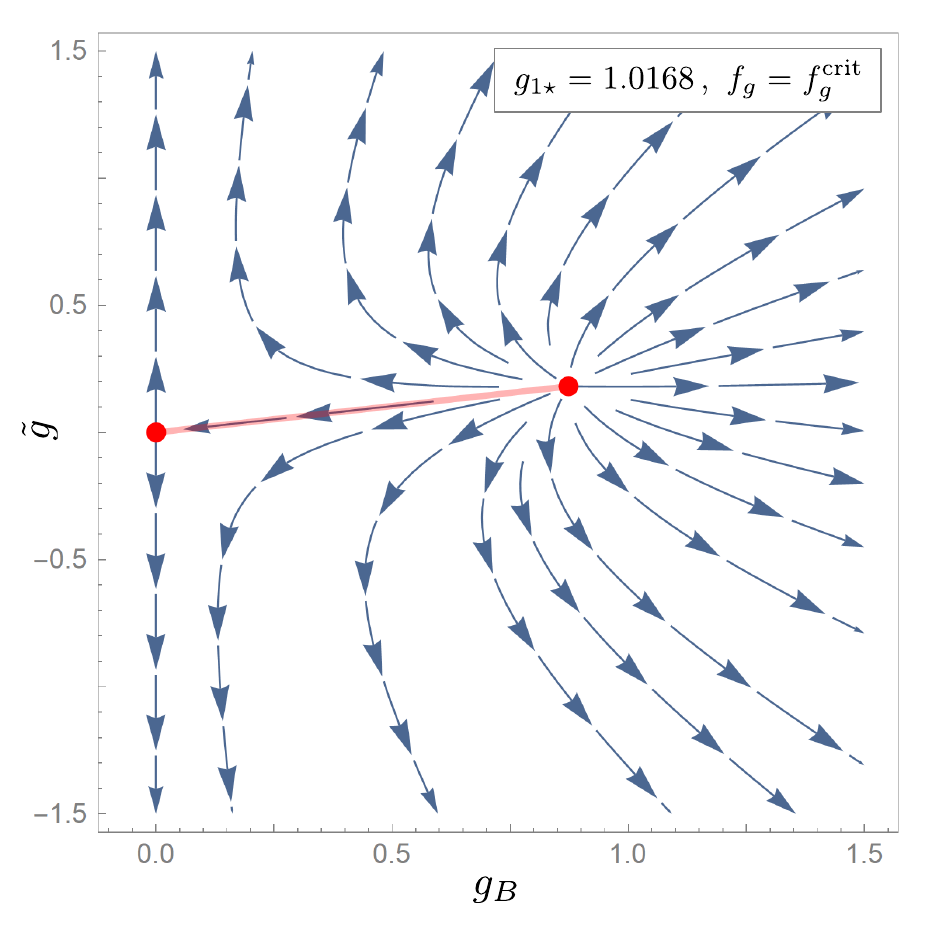}
\caption{UV critical surfaces in the $g_B\tilde{g}$-plane. The interiors of the ellipse and the line are driven to Gaussian fixed points $(g_B,\tilde{g}) = (0,0)$, whereas the ellipse's boundary and the line's right endpoint are non-trivial fixed points. Couplings outside the ellipse and outside the line interval are driven to infinite values.}
\label{fig:1} 
\end{figure}

The remaining $g_1$ fixed point is non-trivial.  In our one-loop approximation, the value $g_{1\star}$ obtained at infinite energy is also the value at the Planck scale, since the $g_1$ 
$\beta$-function vanishes for $\mu \geq M_{\rm Pl}$ due to the choice of $f_g$,
\begin{align}
g_{1\star} = g_1(M_\text{Pl}) = \sqrt{\frac{10}{77}\hat{f}_g} \, .
\label{eq:g1nz}
\end{align}
Since the value of $g_1$ at the Planck scale is fixed by the experimental value at $\mu_0$, $\hat{f}_g$ is then determined
\begin{align}
\hat{f}_{g} = \frac{77}{10} g_{1\star}^2 = \frac{77}{10} g_\text{Planck}^2 \equiv \hat{f}^\text{crit}_g \approx 7.9610 \, .
\end{align}
This corresponds to $f_g \approx 0.05$, consistent in magnitude with typical estimates of $f_g$ appearing in the literature~\cite{Reichert:2019car}. Inserting this critical value $\hat{f}_g^\text{crit}$, one finds the two UV fixed points
\begin{align}
(g_B,\tilde{g}) = (0,0) \, , \quad
(g_B,\tilde{g}) = \left( \sqrt{\tfrac{231 \hat{f}^\text{crit}_g}{2413}}, \sqrt{\tfrac{768 \hat{f}^\text{crit}_g}{185801}}\right) \, .
\label{eq:gbgtnz}
\end{align}
We again plot the $\beta$-functions in the $g_B\tilde{g}$-sector, albeit now for finite $g_{1\star}$, in Fig.~\ref{fig:1}. In this case, we see that the non-trivial $g_B\tilde{g}$ fixed point is connected to the 
Gaussian fixed point by a line. All values that fall onto this critical line (aside from the unstable fixed point at the right end) flow towards that Gaussian fixed point, whereas all values outside this interval are driven to infinite values.  It is 
interesting to observe that the range of $g_B$ coincides with the predicted range in the Gaussian $g_1$ fixed point scenario, whereas the maximum value of $\tilde{g}$ is much smaller.  
Similar to the previous case, the line represents the projection of the UV critical surface into the subspace $g_1=g_{1\star}$ and again implies that the values of $g_B$ and $\tilde{g}$ are not independent.

\begin{figure}[!htb]
\centering
\includegraphics[width=0.48\textwidth]{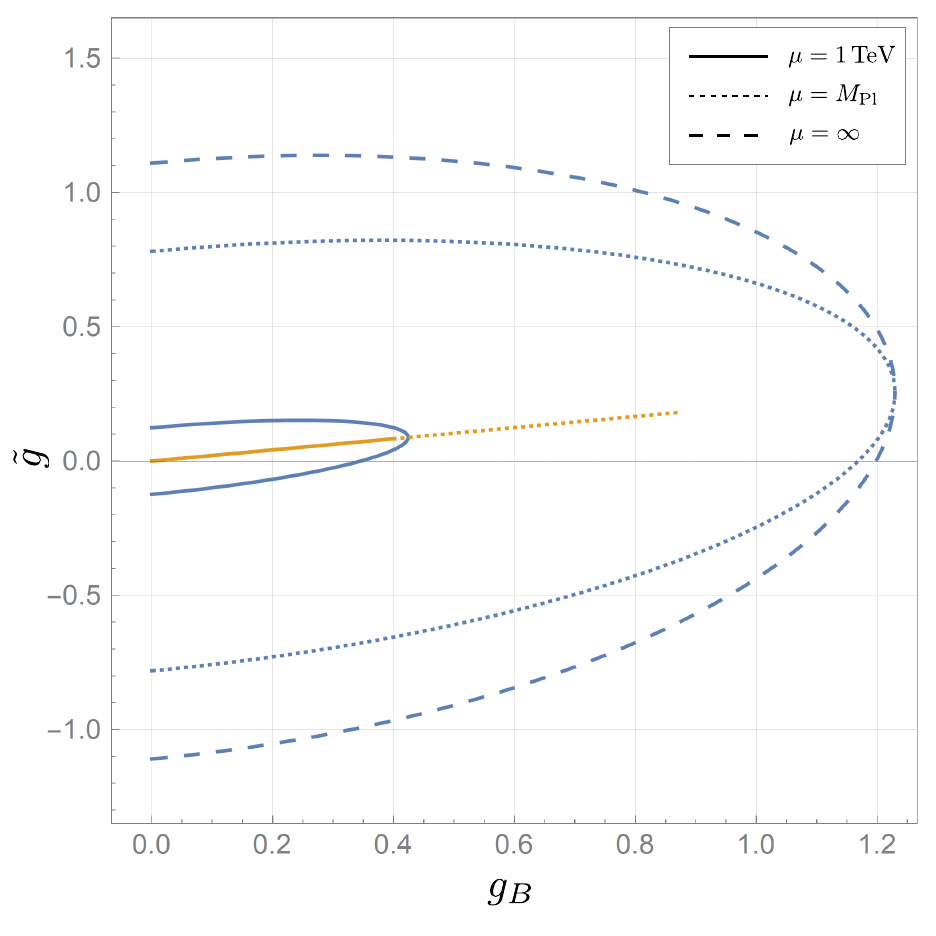}
\includegraphics[width=0.48\textwidth]{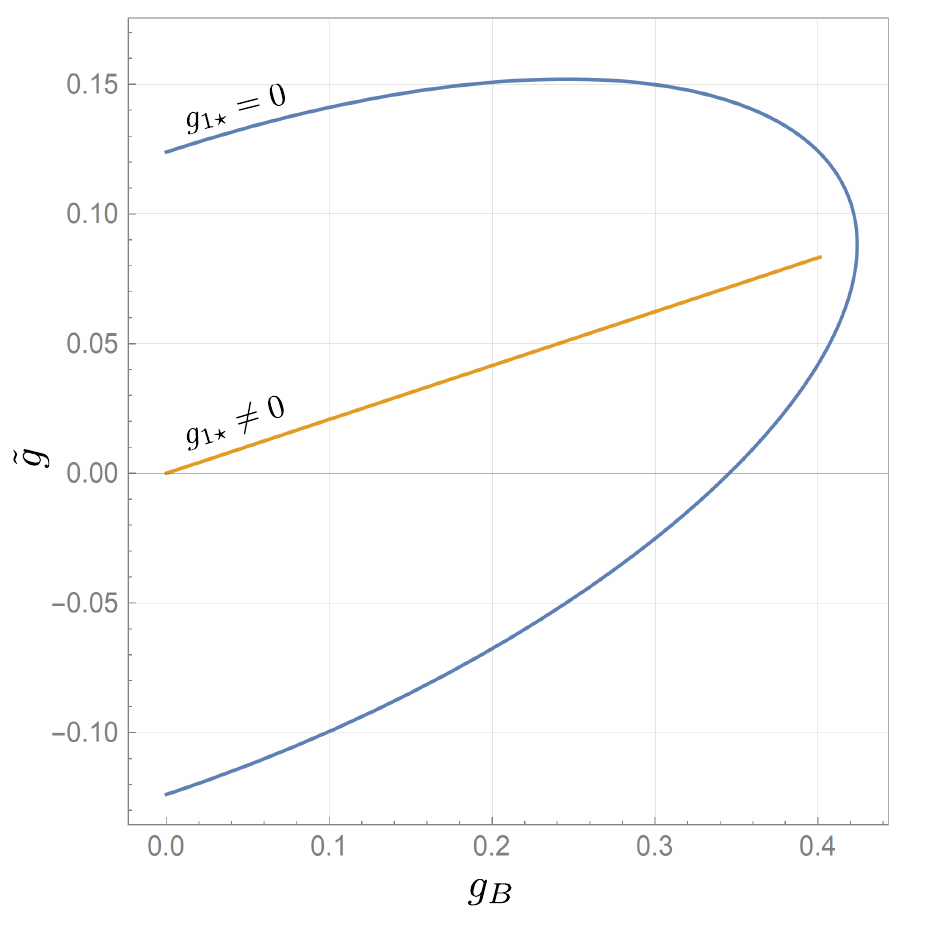}
\caption{Left: UV critical surfaces (dashed line) mapped back to $\mu = M_\text{Pl}$ (dotted line) and to $\mu_0 = 1\,\text{TeV}$ (solid line). The Gaussian $g_1$ fixed point is described by the ellipses
and the interacting $g_1$ fixed point by the line. The requirement that either the boundary of the ellipse or the line is reached leads to a unique relation between $g_B$ and $\tilde{g}$, thereby reducing the degrees of freedom in parameter space of the model. Right: Infrared values in detail.}
\label{fig:2}
\end{figure}

Since the fixed point values of the couplings in Eqs.~(\ref{eq:g1nz})-(\ref{eq:gbgtnz}) are reached at $\mu=M_{\rm Pl}$, it is straightforward to flow these back to our reference scale of $\mu_0 = 1\,\text{TeV}$; see Fig.~\ref{fig:2} for the fixed point coupling values renormalized at $\mu=\infty$, $M_{\rm Pl}$ and 1\,TeV. Again, we find a correlation between the parameters $g_B$ and $\tilde{g}$ at low energy scales.

\subsection{Running couplings and stability of the Higgs sector}
As is well known, the Higgs vacuum of the standard model is metastable. The model under consideration here has extended Higgs, gauge and fermion sectors, which affect the stability analysis and alter this conclusion. The Higgs potential in our model is  given by Eq.~(\ref{eq:thepot}), repeated here for convenience,
\begin{align}
V = -m_H^2 H^\dagger H + \frac{\lambda}{2}(H^\dagger H)^2 - m_\phi^2 \phi^\ast \phi + \frac{\lambda_\phi}{2} (\phi^\ast\phi)^2 + \lambda_m \phi^\ast \phi H^\dagger H \, ,
\label{eq:thepot2}
\end{align}
where $\lambda_m$ couples the two scalars $H$ and $\phi$, and leads to mass mixing after these fields develop vevs.  In unitary gauge,
\begin{align}
H = \frac{1}{\sqrt{2}} \begin{pmatrix}0 \\ v + h \end{pmatrix} \, , \quad \phi = \frac{v_\phi + \varphi}{\sqrt{2}} \, ,
\end{align}
where we have denoted the vevs $v$ and $v_\phi$. Substituting into Eq.~(\ref{eq:thepot2}), one finds the following mass-squared matrix:
\begin{align}
M^2 = \begin{pmatrix} \lambda \, v^2 & \lambda_m v \, v_\phi \\ \lambda_m v \, v_\phi & \lambda_\phi v_\phi^2 \end{pmatrix} \, .
\label{eq:scmm}
\end{align}
The two eigenvalues are 
\begin{align}
m_\pm^2 = \frac12 \left[ \lambda \, v^2 + \lambda_\phi v_\phi^2 \pm \sqrt{\lambda^2v^4 + \lambda_\phi^2 v_\phi^4 - 2 \, v^2v_\phi^2(\lambda\lambda_\phi - 2\lambda_m^2)} \, \right] \, .
\end{align}
As can be seen from the square root, if $\mathfrak{s} \equiv \lambda\lambda_\phi-\lambda_m^2 < 0$ the eigenvalue $m^2_-$ will become negative, indicating that we are no longer at a local minimum.
This can also be verified via the determinant of the mass-squared matrix,
\begin{align}
\det M^2 \equiv m_+^2 m_-^2 = v^2 v_\phi^2(\lambda\lambda_\phi - \lambda_m^2) \equiv v^2v_\phi^2 \mathfrak{s} \, .
\end{align}
Therefore, we require that $\mathfrak{s}>0$ henceforth.  This condition is also sufficient to guarantee stability of the potential at large field amplitudes, since
Eq.~(\ref{eq:thepot2}) can be written as
\begin{equation}
V = -m_H^2 H^\dagger H- m_\phi^2 \phi^\ast \phi+\frac{\lambda}{2}\left[H^\dagger H+\frac{\lambda_m}{\lambda} \phi^\ast \phi \right]^2 + \frac{\mathfrak{s}}{2 \,\lambda} \left[\phi^\ast \phi\right]^2 \,\,\,.
\end{equation}
The last two terms are positive definite when
\begin{equation}
\mathfrak{s}>0 \,\,\,\,\, \mbox{ and } \,\,\,\,\, \lambda>0 \, .
\label{eq:stabcon}
\end{equation}
These inequalities reduce to $\lambda_\phi>0$ and $\lambda>0$ in the case where $\lambda_m=0$, the expected constraints on the quartic terms in two decoupled scalar sectors; for a similar analysis, see Ref.~\cite{Lebedev:2012zw}. In studying the RG evolution of the model, we may now track the sign of $\mathfrak{s}$ and $\lambda$ to confirm that stability of the scalar potential is maintained.   We will see in our subsequent examples that this is the case, and is also consistent with 
the existence of fixed points in the $(\lambda,\lambda_\phi,\lambda_m)$ parameter space that we will explicitly identify.

In the numerical examples that we present in the next two subsections, we adopt the following measured values of the standard model couplings, renormalized at $\mu_0=1$~TeV, also used by Hiller \textit{et al.}~\cite{Hiller:2020fbu}:
\[
\begin{array}{ccccc}
g_1(\mu_0) = 0.46738, &\hspace{1em} & g_2(\mu_0) = 0.63829, & \hspace{1em} & g_3(\mu_0) = 1.05737, \\
\end{array}
\]
\begin{equation}
\begin{array}{ccc}
 y_t(\mu_0) = 0.85322, & \hspace{1em} & y_b(\mu_0)= 0.01388 \, .
\end{array}
\end{equation}
We take the scale of U(1)$_B$ breaking to be $v_\phi=10$~TeV, and require that the lightest scalar mass eigenstate correspond to the Higgs boson, with $m_h=125$~GeV.    To limit
the scope of our following considerations somewhat, we choose
\begin{equation}
\begin{array}{ccc}
\lambda_\phi(\mu_0)=0.2, & \hspace{1em} & \lambda_m(\mu_0) = -0.004 \, .
\end{array}
\end{equation}
We note that the small negative value of $\lambda_m(\mu_0)$ seemed to lead more readily to solutions with the desired vacuum stability.
With the Higgs doublet vev set at $v=246$~GeV, the requirement that we obtain the correct Higgs boson mass then fixes $\lambda(\mu_0)= 0.25828$.  The heavier scalar mass eigenstate will
then have a mass of $4.472$~TeV, heavy enough to not be of immediate phenomenological concern.\footnote{For example, given these choices, the mixing angle that diagonalizes Eq.~(\ref{eq:scmm}) is
${\cal O}(10^{-4})$, compared to the experimental bound from Higgs signal strength measurements that is ${\cal O}(10^{-1})$~\cite{Bause:2021prv}.} Finally we set the Yukawa couplings
\begin{equation}
\kappa_i(\mu_0) = y_i(\mu_0) =  0.1\,\,\,, \,\,\,\,\, i=1\ldots3,
\label{eq:kiyi}
\end{equation}
 and the gravitational correction parameters
\begin{equation}
f_\lambda = f_y = 0.1 \,\, .
\label{eq:flfy}
\end{equation}
The value of the gravitational parameter $f_g$ depends on whether we study the Gaussian $g_1$ fixed point $(f_g>f_g^\text{crit})$, or the interacting one $(f_g=f_g^\text{crit})$.   The values of the remaining couplings and $f_g$ in these two cases are summarized in Table~\ref{table:1}.   Note that the values of $f_\lambda$ and $f_y$ in Eq.~(\ref{eq:flfy}), as well as the value of $f_g$ assumed in the case of the Gaussian $g_1$ fixed point, are roughly comparable in magnitude to that of $f_g$ in the interacting $g_1$ fixed point scenario where the gravitational parameter is determined by the measured value of $g_1$ at low energies.

\begin{table}
\begin{tabular}{cccccccccccc} \hline\hline
case 	&&				&& $g_B$ && $\tilde{g}$ && $m_B$ & \hspace{1em}& $f_g$ \\ \hline
  $g_{1\star} = 0$: && &&  0.3 && 0.14988 && 3\,TeV && 0.1 $(>f_g^\text{crit})$ \\ 
 $g_{1\star} \not= 0$: && && 0.40128 && 0.08338 && 4.01\,TeV && 0.05041 $(=f_g^\text{crit})$ \\ \hline \hline
\end{tabular}
\caption{Remaining model parameters and $Z'$ masses for the two scenarios described in the text, renormalized at the reference scale $\mu_0 = 1\,\text{TeV}$.   The specific values of $g_B$ and 
$\tilde{g}$ in these two examples were selected since they run to fixed point values $g_{B\star}$ and $\tilde{g}_\star$ that are both
nonvanishing.  The value of $f_g$ in the $g_{1\star} \not= 0$ case is set by the requirement 
that the low-energy value of the hypercharge gauge coupling is reproduced.}
\label{table:1}
\end{table}

\begin{figure}[h]
\centering
\includegraphics[width=0.48\textwidth]{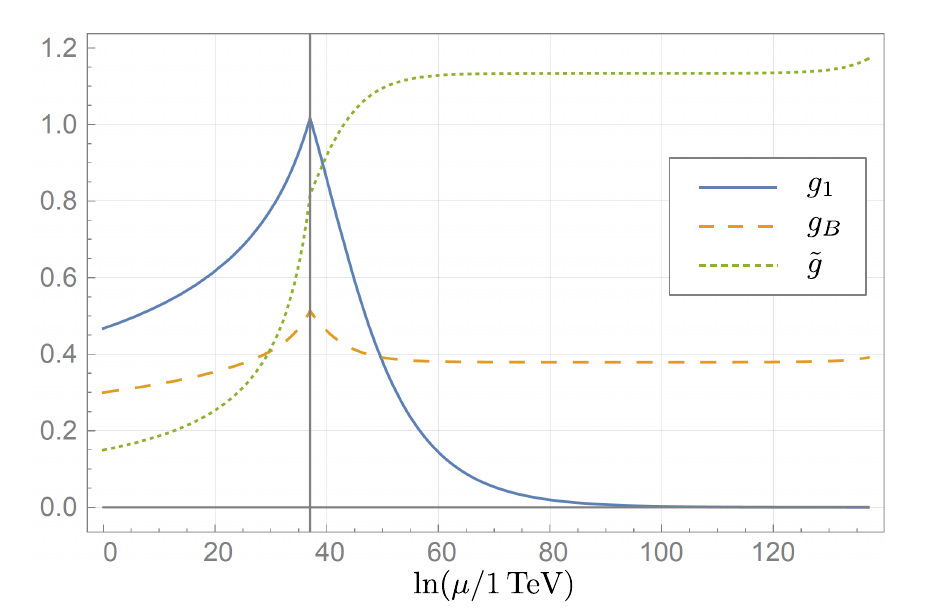}
\includegraphics[width=0.48\textwidth]{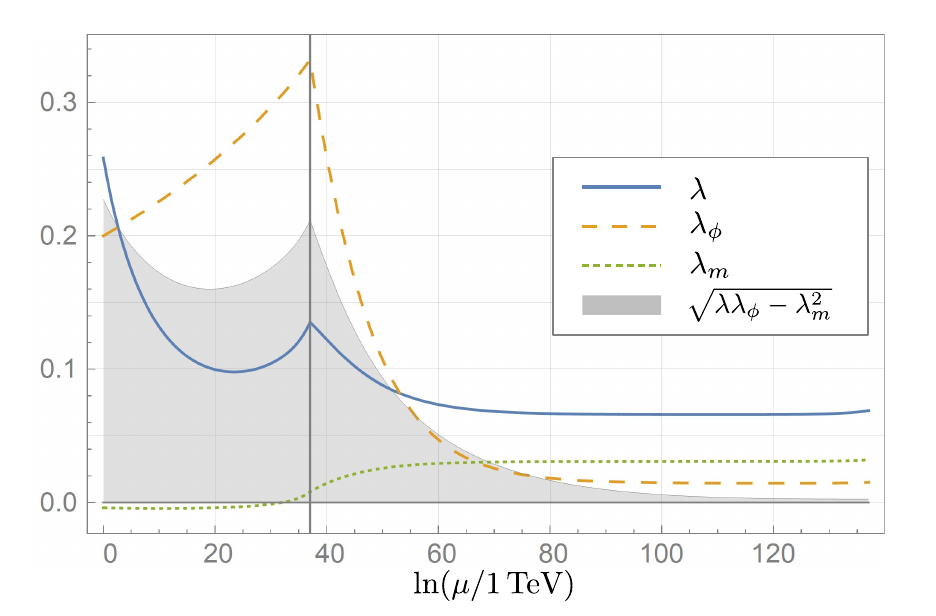}
\includegraphics[width=0.48\textwidth]{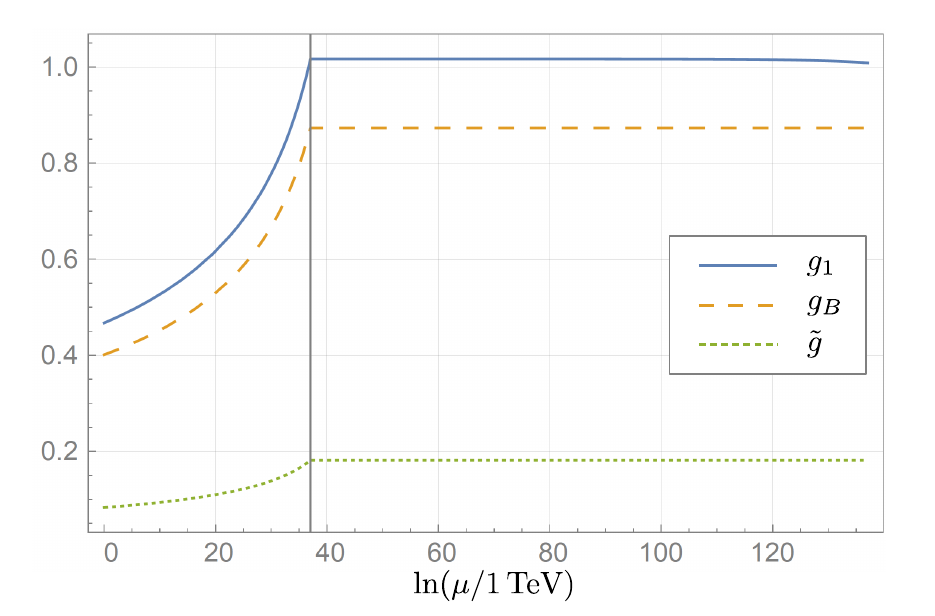}
\includegraphics[width=0.48\textwidth]{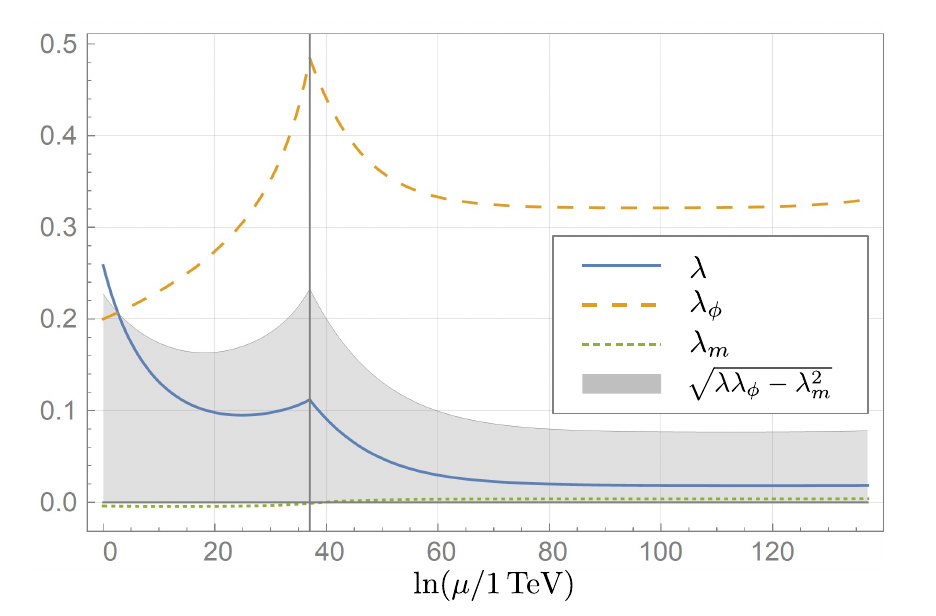}
\caption{Renormalization group flow of the Abelian gauge couplings and the quartic couplings. Top: Gaussian $g_1$ fixed point, bottom: interacting $g_1$ fixed point. In the right plots, the shaded areas highlight the value of $\sqrt{\mathfrak{s}} \equiv \sqrt{\lambda\lambda_\phi-\lambda_m^2}$, where $\mathfrak{s}$, $\lambda$ and $\lambda_\phi$ remain positive  throughout the entire energy range considered here. The quartic couplings attain their numerically expected fixed point values within our numerical resolution, strongly suggesting stability of the Higgs sector up to arbitrarily high energies. The small deviations of the couplings away from their fixed point values visible at the right sides of the plots is an artifact caused by the finite resolution of our numerical approximation and the fact that the fixed points are unstable.  By increasing the resolution of the initial conditions at 1\,TeV, these deviations can be pushed out to arbitrarily high energies. At infinite initial resolution, the couplings would hit their fixed point values exactly.}
\label{fig:3}
\end{figure}

\subsubsection{Gaussian $g_1$ fixed point} \label{subsec:gaussian2}
In this example, we take $f_g = 0.1 > f_g^\text{crit}$ so that we attain a Gaussian fixed point in $g_1$, and choose $g_B(\mu_0) = 0.3$ and $\tilde{g}(\mu_0) = 0.14988$, which assures that these couplings
flow to  a point on the UV ellipse of fixed points discussed earlier and displayed in Fig.~\ref{fig:2}.  We plot the RG flow in the upper two panels of Fig.~\ref{fig:3}. The Higgs sector remains stable throughout, and by making use of the 1-loop $\beta$-functions we find the following fixed points for the quartic couplings:
\begin{align}
\lambda_\star = 0.065871 \, , \quad
\lambda_{\phi\star} = 0.014456 \, , \quad
\lambda_{m\star} = 0.030760 \, .
\end{align}
These values satisfy the stability conditions Eq.~(\ref{eq:stabcon}), and we have confirmed that they are approached in the numerical results presented in Fig.~\ref{fig:3}. 

\subsubsection{Interacting $g_1$ fixed point} \label{subsec:nontrivial2}
Let us now consider $f_g = f_g^\text{crit}$ such that we obtain an interacting $g_1$ fixed point, $g_{1\star} = g_{1}(M_\text{Pl}) = 1.0168$. We further take $g_B(\mu_0) = 0.40128$ and 
$\tilde{g}(\mu_0) = 0.08338$ (the right endpoint of the solid, TeV line in the second panel of Fig.~\ref{fig:2})
to generate a non-zero fixed point for $g_B$ and $\tilde{g}$ as well.  

\begin{figure}[h]
\centering
\includegraphics[width=0.48\textwidth]{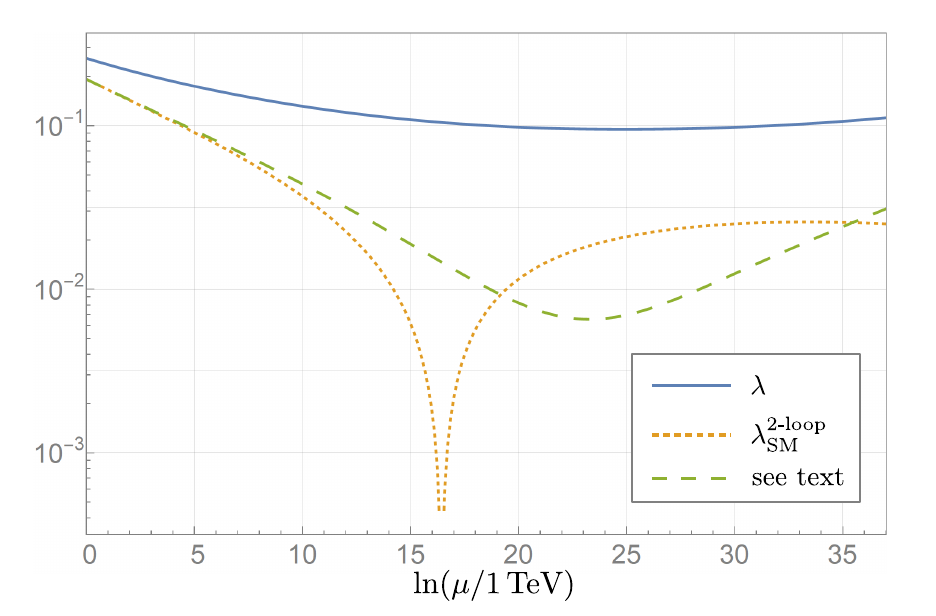}
\includegraphics[width=0.48\textwidth]{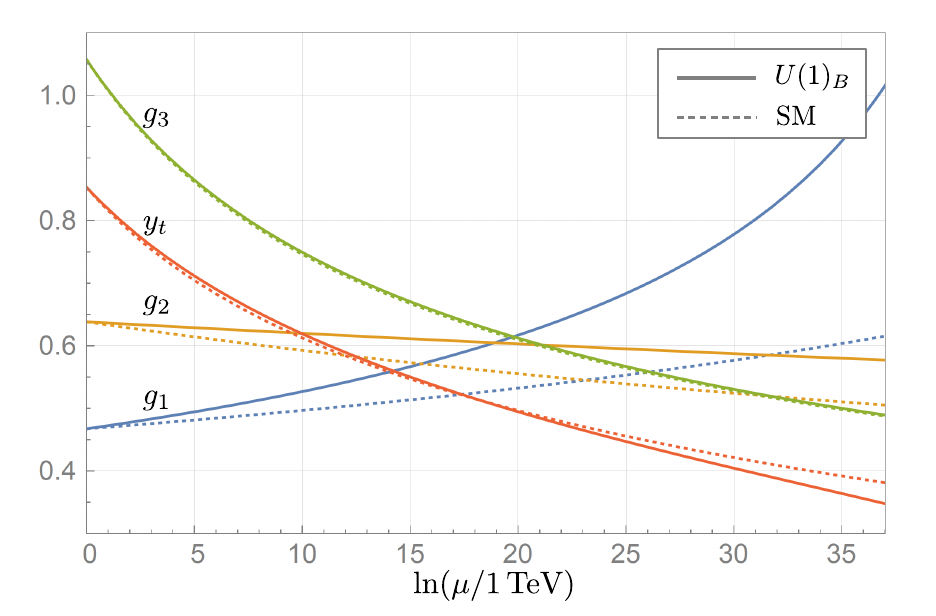}
\caption{Comparison to the standard model.  See the text for discussion.}
\label{fig:4} 
\end{figure}

We plot the corresponding RG flow in the lower two panels of Fig.~\ref{fig:3}. The Higgs sector again remains stable, and we extract the following fixed points:  
\begin{align}
\lambda_\star = 0.018256 \, , \quad
\lambda_{\phi\star} = 0.32076 \, , \quad
\lambda_{m\star} = 0.0037738 \, .
\end{align}
These again are consistent with our stability criteria and agree with our numerical RG flow.   In Fig.~\ref{fig:4}, we show the running of couplings in the present scenario compared to that of the standard model, up to the Planck scale.   The curve for the quartic coupling in the standard model was computed at two loops and assumes the value of $\lambda_{{\rm SM}}$ extracted from the one-loop effective potential, $\lambda_{{\rm SM}}(\mu_0)=0.19234$~\cite{Hiller:2020fbu}; this allows easy comparison with what is typically displayed in the literature.  The dashed line shows what we would find for the coupling $\lambda$ in our model if we were to assume a boundary value identical to that of the standard model curve and also work at two loops.  This illustrates that our model's ability to avoid the metastability of the standard model is a consequence of the new contributions to the $\beta$-functions rather than a different boundary condition at the TeV scale caused by the nonvanishing portal coupling $\lambda_m$.  Analogous plots can be generated for the $g_{1\star}=0$ scenario, but they are qualitatively indistinguishable from those shown in Fig.~\ref{fig:4}, and are not displayed.

\begin{figure}[b]
\centering
\includegraphics[width=0.3\textwidth]{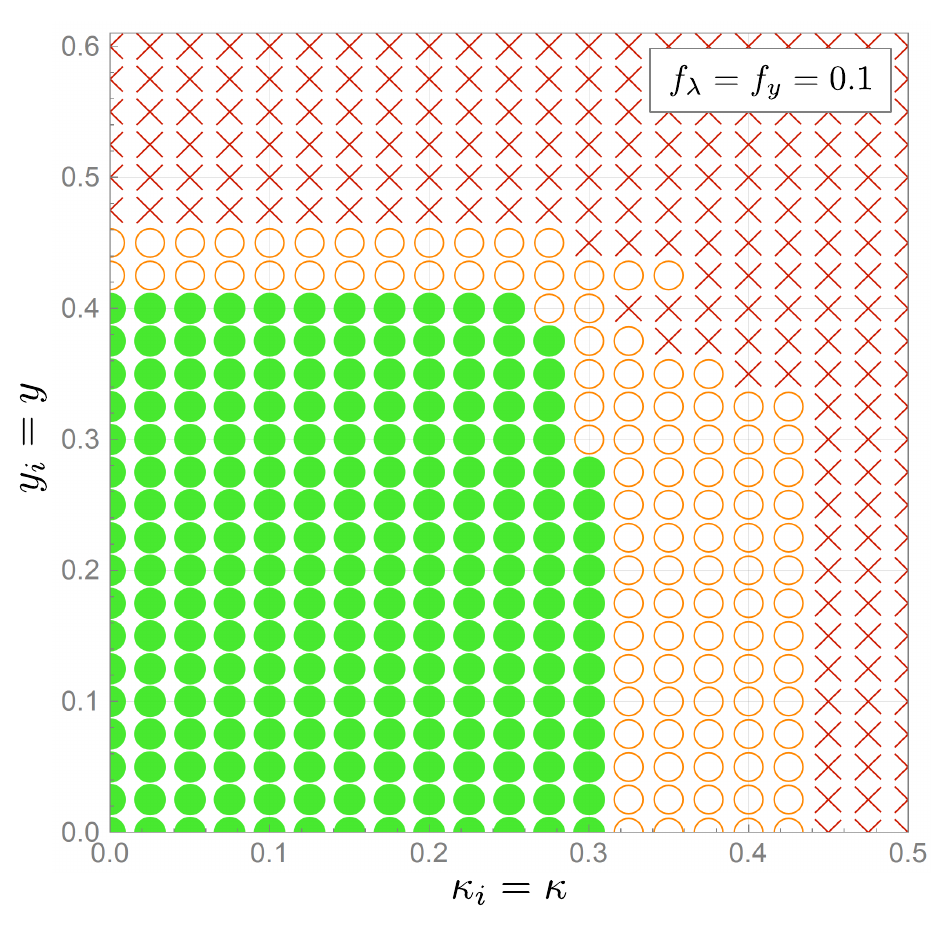}\qquad
\includegraphics[width=0.3\textwidth]{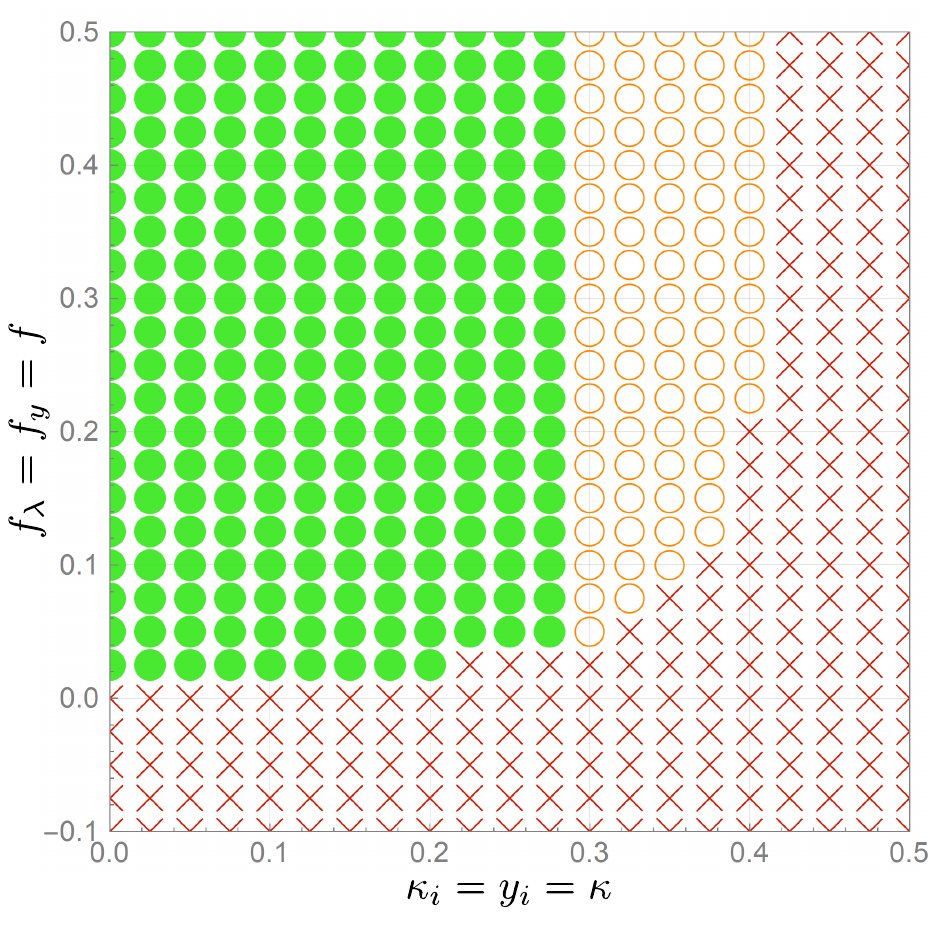}\qquad
\includegraphics[width=0.3\textwidth]{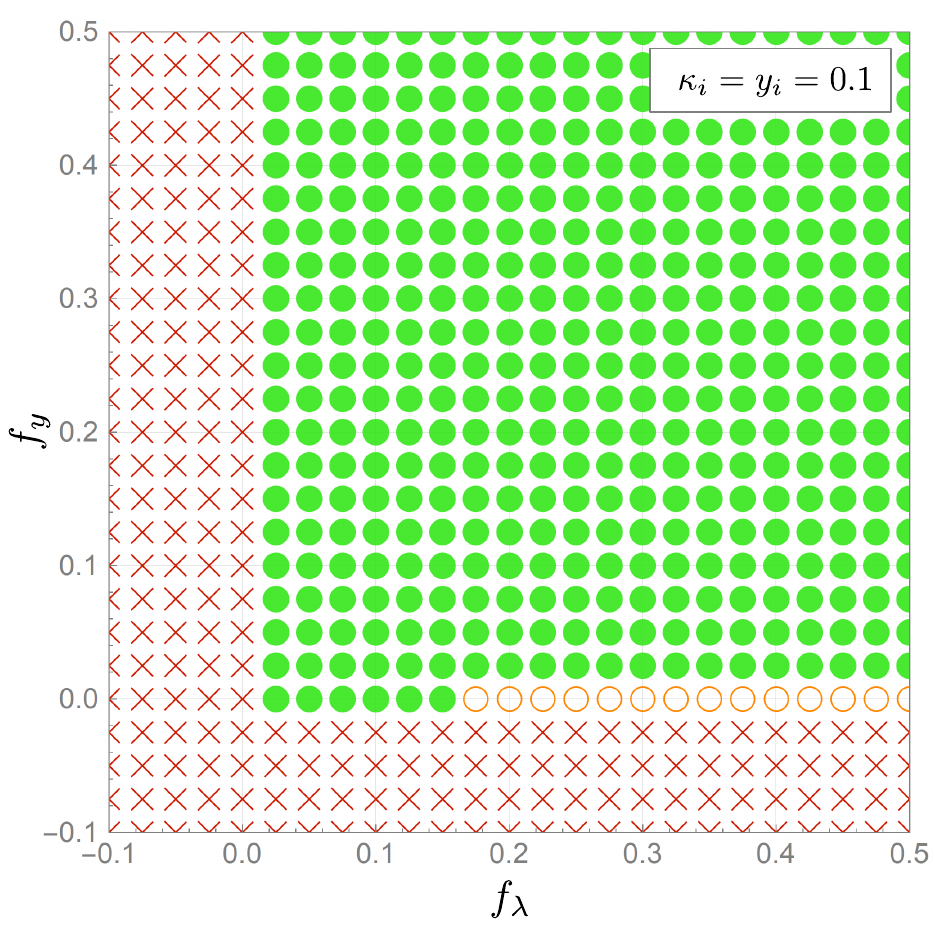}\\
\includegraphics[width=0.3\textwidth]{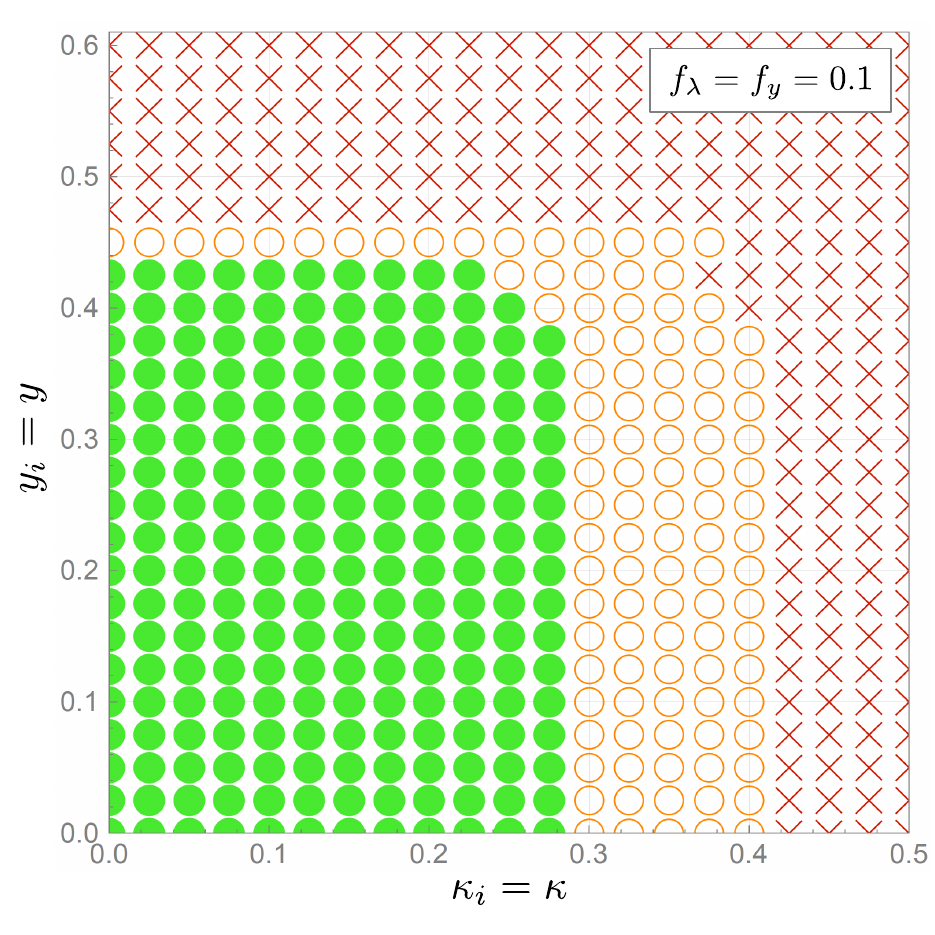}\qquad
\includegraphics[width=0.3 \textwidth]{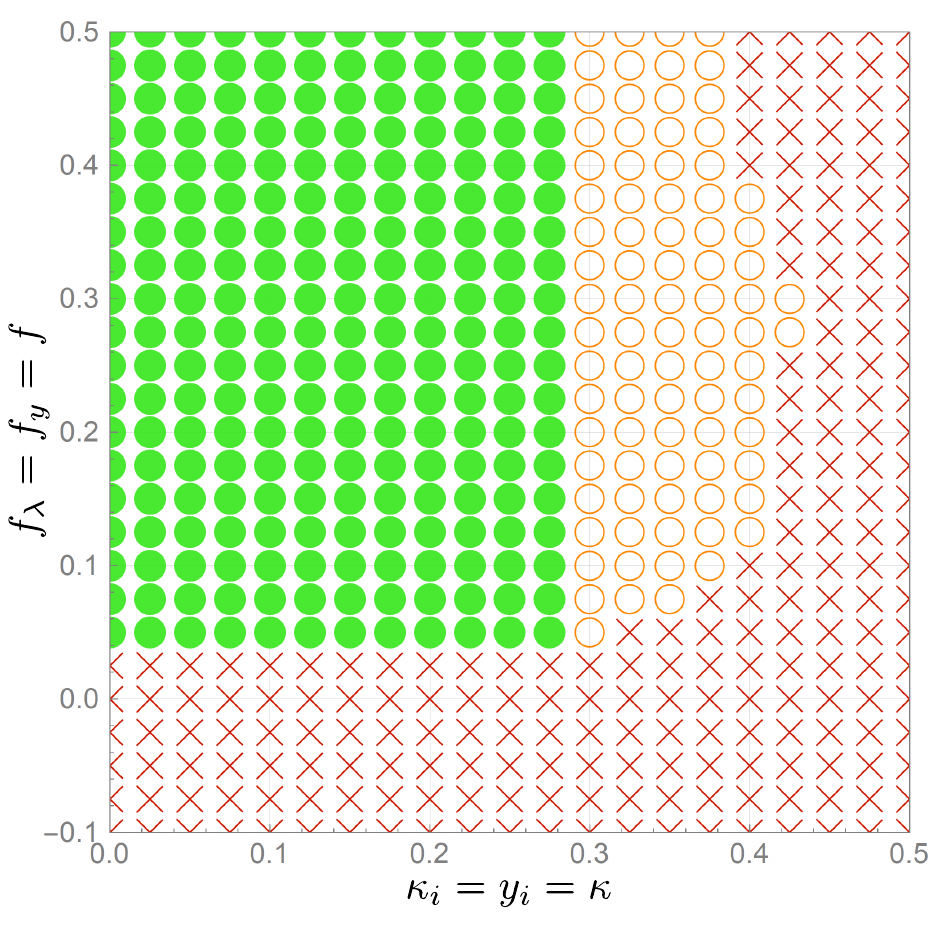}\qquad
\includegraphics[width=0.3\textwidth]{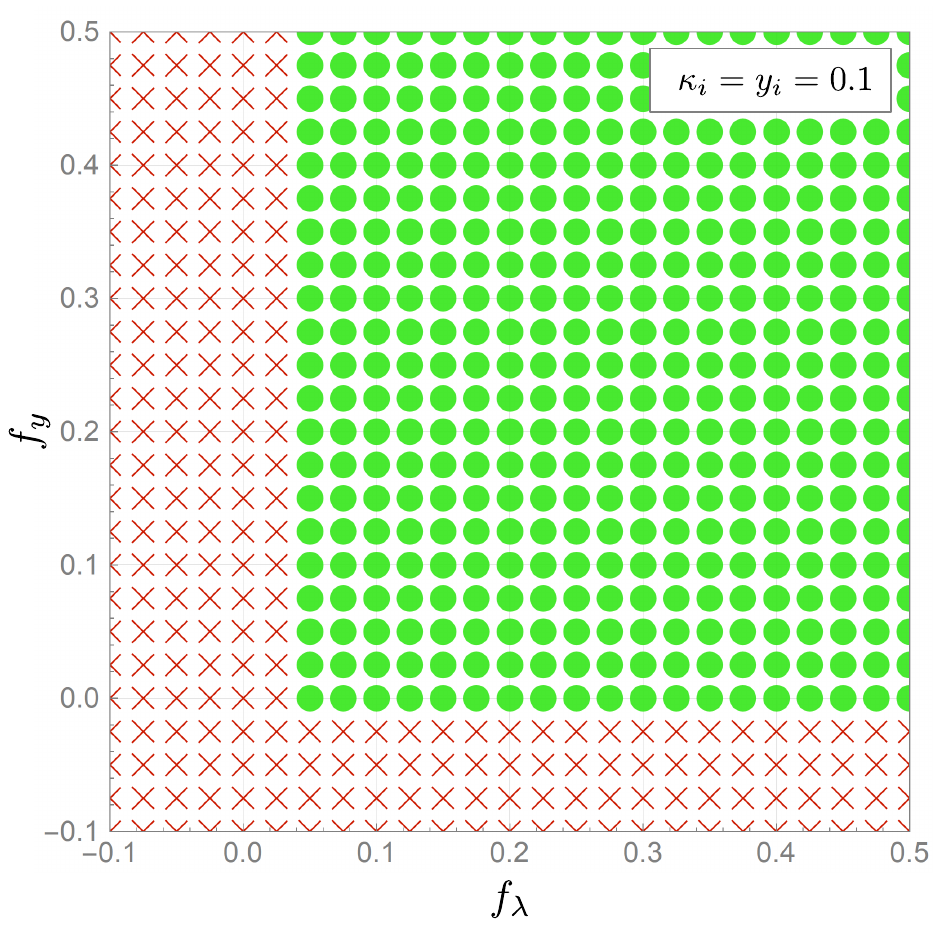}\\
\caption{The effects of varying input parameter values.  The first row corresponds to the Gaussian $g_1$ fixed point, while the second corresponds to the interacting one. The solid circles indicate
viable solutions with perturbative fixed points that satisfy Eq.~(\ref{eq:stabcon}) everywhere, while the crosses represent 
excluded points.  The open circles indicate models with vacuum metastability.}
\label{fig:5} 
\end{figure}

Finally, we note that solutions like those presented in this section can be obtained for other values of the parameters in Eqs.~(\ref{eq:kiyi}) and (\ref{eq:flfy}).  This is illustrated in Fig.~\ref{fig:5}, where we allow these parameters to vary and determine whether viable solutions are obtained.   The solid circles represent solutions where all couplings reach perturbative ultraviolet fixed points and our conditions for vacuum stability are satisfied. Due to the large number of model parameters, there are many possible two-dimensional plots of this type that one could construct;  however, Fig.~\ref{fig:5} is sufficient to demonstrate that Eqs.~(\ref{eq:kiyi}) and (\ref{eq:flfy}) do not represent special choices.
\section{Branching fractions} \label{sec:four}

While it is not the purpose of the present work to engage in an exhaustive phenomenological study of this model (studies of gauged baryon number in a more general context exist in the 
literature~\cite{Carone:1994aa,Carone:1995pu,FileviezPerez:2010gw,Dobrescu:2021vak}), we would like to illustrate in this section how the UV restrictions placed on the parameter space at the TeV scale can lead to 
meaningful predictions of the $Z^\prime$ boson properties.   To do so, we focus on the case of the interacting $g_1$ fixed point, discussion in Secs~\ref{subsec:nontrivial1} and \ref{subsec:nontrivial2}.  In the previous section, we found that the ultraviolet critical surface corresponded to a line segment 
in the $g_B\tilde{g}$-plane which, when run down to the TeV scale, is given by
\begin{equation}
\tilde{g} = \frac{16}{77} \, g_B \,\,\,\,\, \mbox{ for } \,\,\,\,\,  0 \leq g_B \alt 0.4013  \,\,\, .
\label{eq:uvcs}
\end{equation}
Let us consider the implications of this result for branching fractions of the $Z^\prime$ boson.  For a $Z^\prime$ in the multi-TeV mass range, it is a reasonable approximation to neglect the masses of all the standard model
particles (we comment on the effect of including them later).   In this case, the partial decay widths take relatively simple form.   For decays to fermions with $N_c$ colors one has
\begin{equation}
\Gamma (f \bar{f}) = \frac{N_c}{48 \pi} \, \left( C_V^2 + C_A^2 \right) m_B g_B^2 \,\,\, ,
\label{eq:ffbar}
\end{equation}
where the $C_V$ and $C_A$ can be derived from the form of the covariant derivative, Eq.~(\ref{eq:covd}), and we use Eq.~(\ref{eq:uvcs}) 
to eliminate any dependence on $\tilde{g}$.    Numerically, we find that $|C_V|$ is given by $0.8008$, $0.6398$, $0.2414$, and $0.0805$ for the up-type quarks, down-type quarks, charged leptons, and neutrinos, respectively;  the $|C_A|$ are each $0.0805$.   This is sufficient to
determine the partial width to dijets (including all quarks except the top), and to charged dileptons.   We also take into account that there are decays to $W^+ W^-$: this is easiest to 
compute in the original basis where the coupling $\epsilon$ is present in the gauge boson kinetic terms and is treated here as a perturbative interaction.  In this case, at lowest order in $\epsilon$, we find (with help from Feyn Calc~\cite{feyncalc})
\begin{equation}
\Gamma(W^+ W^-) = \frac{\alpha \cos\theta_w^2 \epsilon^2}{12} m_B \frac{y^4}{(1-y^2)^2} \sqrt{1-\frac{1}{x^2}} \left( 4x^4 + 16x^2 - \frac{3}{x^2} - 17 \right) \,\,\, ,
\end{equation}
where $x=\frac{m_B}{2 m_W}$, $y=m_Z/m_B$ and $\theta_w$ is the weak mixing angle.  Using $m_W =m_Z \cos \theta_w$ and assuming
$m_B \gg m_Z$, one may show that this result approaches
\begin{equation} 
\Gamma(W^+ W^-) \approx \frac{4}{29645 \pi} m_B g_B^2 \,\,\, ,
\label{eq:wpwm}
\end{equation}
provided the kinetic mixing is small.  The consequence of Eq.~(\ref{eq:uvcs}) is that Eqs.~(\ref{eq:ffbar}) and (\ref{eq:wpwm}) are proportional to $m_B g_B^2$, which implies that the $Z^\prime$ branching fractions are approximately fixed provided that the $Z^\prime$ boson is sufficiently heavy and that we live within the range $0 \leq g_B \alt 0.4013$.  We find that
\begin{align}
\begin{split}
& \text{BF}(Z^\prime \rightarrow {\rm jets}) = 77.8 \, \%  \\
& \text{BF}(Z^\prime \rightarrow t \overline{t}) = 19.8 \, \% \\
& \text{BF}(Z^\prime \rightarrow {\ell^+\ell^-}) = 2.0 \, \% \\
& \text{BF}(Z^\prime \rightarrow W^+ W^-) = 0.1 \, \%
\end{split}
\end{align}
with the remainder going to invisible decays (\textit{i.e.}, neutrinos).  For example, $m_B= 3$~TeV and $g_B=0.3$ is a choice that satisfies our 
assumptions and is consistent with current experimental bounds. For this point in model parameter space, we have checked that the effect of 
including final state particle masses, including that of the top quark, affects the branching fractions shown above only at the next decimal place.   
LHC searches for new resonances decaying to dijets  allow the $Z^\prime$ of the present model for $m_B=3$~TeV and $g_B=0.3$ (see Fig.~3 in 
Ref.~\cite{Dobrescu:2021vak}, where the value of the coupling to quarks would be 0.6 in their conventions, well within their allowed region.).  
Moreover, Eq.~(\ref{eq:uvcs}) implies the value $\tilde{g}=0.0623$, corresponding to the kinetic mixing parameter $\epsilon=0.1321$, consistent with the kinetic mixing bounds in Ref.~\cite{Hook:2010tw} for a $3$~TeV $Z^\prime$. One might expect the model to provide similar predictivity for heavier $Z^\prime$ bosons which will be less constrained by current experimental bounds.

\section{Conclusions} \label{sec:conc}

We have considered a model with gauged baryon number that is asymptotically safe due to gravitational corrections introduced above the Planck scale. Three generations of vector-like fermions, which are chiral under the baryon gauge symmetry, cancel the gauge and gravitational anomalies in the theory. The baryon number gauge symmetry is spontaneously broken after a new complex scalar field obtains a non-zero vacuum expectation value. By requiring that the couplings flow to asymptotic fixed points, we restrict the parameter space of the model, in some cases relating the kinetic mixing to the baryon gauge coupling. This allows us to predict measurable quantities, such as branching fractions, of the $Z^\prime$ boson at future collider experiments.

The ultraviolet behavior of the model depends on the size of the gravitational corrections. For sufficiently large gravitational correction terms in the RGEs, there exists either a Gaussian fixed point or an interacting fixed point for the GUT-normalized hypercharge coupling, $g_1$. For each of these cases, we examined the parameter space of the baryon gauge coupling and the kinetic mixing,  $(g_B,\tilde{g})$, to find the subspace that flows to ultraviolet fixed points, thereby defining the UV critical surface. In the Gaussian case, there is a stable trivial fixed point for $(g_{B\star},\tilde{g}_\star)=(0,0)$ and an unstable ellipse of fixed points whose size is determined by the magnitude of the gravitational correction terms in the RGEs. In the interacting case, one finds a trivial fixed point as well as an unstable non-trivial fixed point.  Any value on the line connecting these fixed points flows to the trivial fixed point.  Any values off of this line or outside of the ellipse flow to infinite values and correspond to unphysical theories.

In the Higgs sector of our model, we stated the conditions on the quartic couplings for vacuum stability and confirmed that they are satisfied under the RG flow. After fixing values for the couplings at the TeV scale and running them up to fixed points, we included gravitational effects on the trans-Planckian RG flow.  Within our stated approximations, we obtained numerical evidence suggesting that the Higgs sector retains its vacuum stability to arbitrarily high energies, with all of its couplings approaching nontrivial fixed points.

We briefly examine the phenomenology of the model by determining the restrictions on the parameter space at the TeV scale imposed by
asymptotic safety.  Considering the case where the $g_1$ fixed point is nonzero, we were able to predict values for the branching fractions of the $Z^\prime$ boson into jets, $t\bar{t}$, charged dileptons, and $W^+W^-$, from the relationship between the kinetic mixing and baryon gauge coupling.  To good approximation, the partial decay widths are proportional to $m_B g_B^2$, when $m_B \gg m_Z$, which implies that the branching fractions are all fixed if the $Z^\prime$ boson is sufficiently massive.

In future work, we look forward to exploring other aspects of the phenomenology of this asymptotically safe gauged baryon number model, including how the heavy vector-like leptons affect the muon $g-2$ and how a viable dark matter candidate may be included.

\begin{acknowledgments} 
We thank Gudrun Hiller for useful clarifications regarding their work, and Marc Sher for valuable comments on the manuscript. We thank the NSF for support under Grants PHY-1819575 and PHY-2112460.
\end{acknowledgments}

\appendix

\section{Renormalization group equations at one loop}
\label{app}

Definition:
\begin{align}
\beta(c) = \sum\limits_{\ell=1}^\infty \frac{1}{(4\pi)^{2\ell}} \beta^{(\ell)}(c)
\end{align}

Gauge couplings:
\begin{align}
\beta^{(1)}(g_1) &= +\frac{77}{10} g_1^{3} \, , \\
\beta^{(1)}(g_B) &= +11 g_B^{3} + \frac{77}{6} g_B \tilde{g}^{2} -  \frac{16}{3} g_B^{2} \tilde{g} \, , \\
\beta^{(1)}(\tilde{g}) &= -\frac{16}{5} g_1^{2} g_B -  \frac{16}{3} g_B \tilde{g}^{2} + \frac{77}{5} g_1^{2} \tilde{g} + 11 g_B^{2} \tilde{g} + \frac{77}{6} \tilde{g}^{3} \, , \\
\beta^{(1)}(g_2) &= -\frac{7}{6} g_2^{3} \, , \\
\beta^{(1)}(g_3) &= -7 g_3^{3} \, .
\end{align}
Yukawas ($\kappa^2 \equiv \kappa_1^{2} + \kappa_2^{2} + \kappa_3^{2}, \,y^2 \equiv 2 y_1^{2} +  y_2^{2} + y_3^{2}$):
\begin{align}
\beta^{(1)}(y_t) &= \left(\frac{9}{2} y_t^{2} + \frac{3}{2} y_b^{2} + 3 \kappa^2 - \frac{17}{20} g_1^{2} - \frac{9}{4} g_2^{2} - 8 g_3^{2} - \frac{2}{3} g_B^{2} - \frac{5}{3} g_B \tilde{g} - \frac{17}{12} \tilde{g}^{2} \right) y_t \, , \\
\beta^{(1)}(y_b) &= \left(\frac{3}{2} y_t^{2} + \frac{9}{2} y_b^{2} + 3\kappa^2 - \frac{1}{4} g_1^{2} - \frac{9}{4} g_2^{2} - 8 g_3^{2} - \frac{2}{3} g_B^{2} + \frac{1}{3} g_B \tilde{g} - \frac{5}{12} \tilde{g}^{2} \right) y_b \, , \\
\beta^{(1)}(y_1) &= \left( 3 y^2 + y_1^2 + \frac{1}{2} \kappa_1^{2} -  \frac{9}{10} g_1^{2} -  \frac{9}{2} g_2^{2} - 3 g_B^{2} + 3 g_B \tilde{g} - \frac{3}{2} \tilde{g}^{2} \right) y_1 \, , \\
\beta^{(1)}(y_2) &= \left( 3 y^2 + y_2^2 + \kappa_2^{2} -  \frac{18}{5} g_1^{2} - 3 g_B^{2} + 6 g_B \tilde{g} - 6 \tilde{g}^{2} \right) y_2 \, , \\
\beta^{(1)}(y_3) &= \bigg( 3 y^2 + y_3^2 + \kappa_3^{2} - 3 g_B^{2} \bigg) y_3 \, , \\
\beta^{(1)}(\kappa_1) &= \left( 3 y_t^{2} + 3 y_b^{2} + \frac{1}{2} y_1^{2} + \frac{9}{2} \kappa_1^{2} + 3 \kappa_2^{2} + 3 \kappa_3^{2} - \frac{9}{4} g_1^{2} - \frac{9}{4} g_2^{2} - \frac{15}{4} \tilde{g}^{2} \right) \kappa_1 \, , \\
\beta^{(1)}(\kappa_2) &= \left( 3 y_t^{2} + 3 y_b^{2} + \frac{1}{2} y_2^{2} + 3 \kappa_1^{2} + \frac{9}{2} \kappa_2^{2} + \frac{3}{2} \kappa_3^{2} - \frac{9}{4} g_1^{2} - \frac{9}{4} g_2^{2} - \frac{15}{4} \tilde{g}^{2} \right) \kappa_2 \, , \\
\beta^{(1)}(\kappa_3) &= \left( 3 y_t^{2} + 3 y_b^{2} + \frac{1}{2} y_3^{2} + 3 \kappa_1^{2} + \frac{3}{2} \kappa_2^{2} + \frac{9}{2} \kappa_3^{2} - \frac{9}{20} g_1^{2} - \frac{9}{4} g_2^{2} - \frac{3}{4} \tilde{g}^{2} \right) \kappa_3 \, .
\end{align}
Quartic couplings ($\kappa^2 \equiv \kappa_1^{2} + \kappa_2^{2} + \kappa_3^{2}$, $K^4 \equiv \kappa_1^{4} + \kappa_2^{4} + \kappa_3^{4}$, 
$y^2 \equiv 2 y_1^{2} +  y_2^{2} + y_3^{2}$,  $Y^4 \equiv 2 y_1^{4} +  y_2^{4} + y_3^{4}$):
\begin{align}
\beta^{(1)}(\lambda) &= +12 \lambda^{2} + 2 \lambda_m^{2} -  \frac{9}{5} g_1^{2} \lambda - 9 g_2^{2} \lambda - 3 \tilde{g}^{2} \lambda \nonumber \\
&\hspace{13pt}+ \frac{27}{100} g_1^{4} + \frac{9}{10} g_1^{2} g_2^{2} + \frac{9}{10} g_1^{2} \tilde{g}^{2} + \frac{9}{4} g_2^{4} + \frac{3}{2} g_2^{2} \tilde{g}^{2} + \frac{3}{4} \tilde{g}^{4} \nonumber \\
&\hspace{13pt}+ 12 \lambda \left( y_t^{2} + y_b^{2} + \kappa^2 \right) - 12 \left( y_t^{4} + y_b^{4} + K^4 \right) \, , \\
\beta^{(1)}(\lambda_\phi) &= +10 \lambda_\phi^{2} + 4 \lambda_m^{2} - 12 g_B^{2} \lambda_\phi + 12 g_B^{4} + 12 \lambda_\phi y^2 - 12\, Y^4 \, , \\
\beta^{(1)}(\lambda_m) &= \Big[ 6 \lambda + 4 \lambda_\phi + 4 \lambda_m - \frac{9}{10} g_1^{2} - \frac{9}{2} g_2^{2} - 6 g_B^{2}  -  \frac{3}{2} \tilde{g}^{2}   + 6 ( y_t^{2} + y_b^{2} + y^2 + \kappa^2 ) \Big] \lambda_m \nonumber \\
&\hspace{13pt} + 3 g_B^{2} \tilde{g}^{2} - 12 (\kappa_1^{2} y_1^{2} + \kappa_2^{2} y_2^{2} + \kappa_3^{2} y_3^{2} ) \, .
\end{align}

\end{document}